\documentclass[aps, prd, reprint, twocolumn,
    amsmath, amssymb,
    floatfix, nofootinbib, preprintnumbers, superscriptaddress, 
]{revtex4-2}

\usepackage{graphicx}
    \graphicspath{{fig/}}
\usepackage{orcidlink}

\hypersetup{
    pdfnewwindow=true,
    colorlinks=true,
    linkcolor=blue,
    citecolor=blue,
    filecolor=blue,
    urlcolor=blue
}

\begin{document}

\preprint{CETUP-2025-002}

\title{
    \emph{No Hiding in the Dark:} \texorpdfstring{\\}{}
    Cosmological Bounds on Heavy Neutral Leptons with Dark Decay Channels
}

\author{P. S. Bhupal Dev\orcidlink{0000-0003-4655-2866}}
\email{bdev@wustl.edu}
\affiliation{Department of Physics and McDonnell Center for the Space Sciences, Washington University, St.~Louis, MO 63130, USA}
\affiliation{PRISMA$^+$ Cluster of Excellence \& Mainz Institute for Theoretical Physics, Johannes Gutenberg-Universität Mainz, 55099 Mainz, Germany}

\author{Quan-feng Wu\orcidlink{0000-0002-5716-5266}}
\email{wuquanfeng@ihep.ac.cn}
\affiliation{Institute of High Energy Physics, Chinese Academy of Sciences, Beijing 100049, China}
\affiliation{Kaiping Neutrino Research Center, Kaiping 529386, China}

\author{Xun-Jie Xu\orcidlink{0000-0003-3181-1386}}
\email{xuxj@ihep.ac.cn}
\affiliation{Institute of High Energy Physics, Chinese Academy of Sciences, Beijing 100049, China}

\begin{abstract}
    Heavy neutral leptons (HNLs) are well-motivated new physics candidates.
    The mixing of sub-GeV HNLs with active neutrinos is severely constrained by cosmology.
    In particular, the success of Big Bang Nucleosynthesis (BBN) requires the HNL lifetime to be shorter than about 0.02 sec if they were in thermal equilibrium, thus excluding a wide range of mixing angles accessible to terrestrial experiments.
    In order to justify the laboratory searches in this cosmologically-forbidden region, it is often argued that adding new dark sector decay modes of HNLs can evade the stringent BBN constraint.
    Here we rule out this possibility and show that, contrary to the naive expectation, HNLs with significant dark decay modes actually lead to stronger cosmological bounds.
    This is mainly because of the increase in the extra radiation energy density in the Universe around the BBN epoch, which causes observable effects in the primordial helium fraction and $\Delta N_{\rm eff}$.
    Our result has major implications for laboratory searches of HNLs.
\end{abstract}

\maketitle

\section{Introduction}

The origin of neutrino mass is a fundamental open question.
Among the many theoretical possibilities, perhaps the simplest is to extend the Standard Model (SM) particle content by adding right-handed neutrinos, popularly known as heavy neutral leptons (HNLs), thus generating the Dirac mass ($m_D$) for neutrinos after electroweak symmetry breaking.
Moreover, the HNLs, being SM-singlets, can also acquire a Majorana mass ($m_N$), which leads to the famous seesaw relation for explaining the light neutrino masses: $m_\nu \simeq -m_D m_N^{-1}m_D^T$ \cite{Minkowski:1977sc, Mohapatra:1979ia, Yanagida:1979as, Gell-Mann:1979vob}.
Irrespective of the origin of the Majorana mass term and its ultraviolet completion, even the minimal $\nu$SM scenario with just SM+HNLs can not only explain neutrino masses, but also account for the observed dark matter and baryon asymmetry of the Universe \cite{Asaka:2005pn, Boyarsky:2009ix}, thus making it an elegant candidate for beyond-the-SM (BSM) physics.

In the $\nu$SM, the HNLs interact with the SM sector only via their mixing with the active neutrinos $\nu_\alpha$ (with $\alpha=e,\mu,\tau$) given by $U_\alpha\simeq m_Dm_N^{-1}$.
The seesaw relation predicts
\begin{equation}
    U_{\alpha}^{2}\simeq \frac{m_{\nu}}{m_N}\ \ (\text{seesaw line})\thinspace,
    \label{eq:seesaw}
\end{equation}
where $m_{\nu} \sim 0.01-0.1$ eV is the neutrino mass scale needed to explain the solar and atmospheric mass-squared differences, respectively.
Note that this relation is strictly valid only in the single-HNL limit, while the general case with two or more HNLs \cite{Kersten:2007vk, He:2009ua, Adhikari:2010yt, Ibarra:2010xw}, or specific model constructions \cite{Mohapatra:1986bd, Malinsky:2005bi, Mitra:2011qr, Dev:2012sg, Lee:2013htl, Chauhan:2020mgv}, can yield significantly larger mixing, thus allowing $m_N$ and $U_\alpha$ to be essentially treated as free parameters.
Nevertheless, Eq.~\eqref{eq:seesaw} offers a theoretically well-motivated target for experimental searches.

A summary of the existing constraints in the HNL mass-mixing plane can be found in Refs.~\cite{Bolton:2019pcu, Abdullahi:2022jlv, Fernandez-Martinez:2023phj}.
For $m_N\lesssim {\cal O}$(GeV), the most stringent constraint on the HNL mixing comes from Big Bang Nucleosynthesis (BBN), which requires that the HNL lifetime must be shorter than $\sim 0.02-0.1$ sec \cite{Sabti:2020yrt, Boyarsky:2020dzc, Chen:2024cla, Akita:2024nam, Akita:2024ork,Ovchynnikov:2024xyd}.
In this paper, we explore the possibility of whether this lifetime constraint can be evaded or at least relaxed in the presence of additional HNL decays to dark sector particles.
We show that, contrary to the naive expectation, just making the HNL more short-lived by dominantly decaying to the dark sector does not actually relax the cosmological bound, but makes it stronger.
This comes from a combination of the BBN constraints on the primordial abundance of light elements and the Cosmic Microwave Background (CMB) constraint on the effective number of relativistic degrees of freedom.
This conclusion is qualitatively expected from energy conservation, since the shortened lifetime generally implies that the energy of HNLs is dominantly transferred to the dark sector, so its impact on the Hubble expansion cannot be hidden.
Nonetheless, a quantitative analysis is still required to evaluate this impact, as we carry out in this paper.

\section{HNLs in the early Universe}

In this work, we are mainly interested in the MeV--GeV scale HNLs and experimentally-relevant mixing above the seesaw line.
In this regime, the mixing yields a sufficiently high interaction rate of HNLs with the SM thermal bath and thus brings the HNLs into thermal equilibrium at an early epoch.
As the Universe cools down to the temperature
\begin{equation}
    T_{i}\simeq\frac{1\ \text{MeV}}{U_{\alpha}^{2/3}}\thinspace,
    \label{eq:Tfo}
\end{equation}
HNLs freeze out from the thermal bath while being relativistic (for the allowed values of mixing), similar to neutrino decoupling in the SM.
Throughout this paper, we use the subscript ``$i$'' to indicate the moment of $T=T_i$.
After decoupling, HNLs start free-streaming in the Universe and eventually decay when the age of the Universe is comparable to their lifetime $\tau_{N}$.

The age of the Universe during the radiation-dominated era can be estimated using $t\approx1/(2H)$, where $H$ is the Hubble parameter.
Using $H\approx1.66g_{\star}^{1/2}T^{2}/m_{{\rm pl}}$, where $g_{\star}$ is the effective number of degrees of freedom, $T$ is the temperature of the Universe, and $m_{{\rm pl}}\approx1.22\times10^{19}$ GeV is the Planck mass, one obtains\footnote{
    Note that in the presence of HNLs, Eq.~\eqref{eq:t-T} could be modified significantly, in particular when HNLs dominate the total energy.
    In this work, Eq.~\eqref{eq:t-T} is only used for qualitative discussions and not used in quantitative calculations.
}
\begin{equation}
    t\approx0.74\ \text{sec}\cdot\left(\frac{\text{MeV}}{T}\right)^{2}\left(\frac{10.75}{g_{\star}}\right)^{1/2}\thinspace.
    \label{eq:t-T}
\end{equation}
At $T\approx2.7$ MeV, the age of the Universe is $0.1$ sec.
As the Universe evolves beyond this point, the SM thermal bath undergoes several crucial processes, including (i) neutrino decoupling, (ii) freeze-out of the neutron-to-proton ratio, and (iii) $e^{+}e^{-}$ annihilation.
All of these processes occur at around the MeV scale and, if influenced by new physics, may alter standard cosmological predictions of BBN and CMB observables such as the helium abundance ($Y_{P}$) and the effective number of relativistic species ($N_{{\rm eff}}$).
In order to keep the successful predictions of the standard cosmology unaffected, HNLs must decay before these crucial processes start to take effect, i.e., before the Universe cools down to the MeV scale.
Consequently, the cosmological bound on HNLs in the freeze-out regime roughly corresponds to an upper bound on their lifetime $\tau_{N}$.
For instance, $\tau_{N}\lesssim0.1\ \text{sec}$ was used in earlier studies \cite{Dolgov:2000jw, Boyarsky:2009ix, Ruchayskiy:2012si} as a conservative estimate of the BBN bound.
More recent studies performed a dedicated analysis of the HNL's influence on the MeV-scale Universe and obtained lower values of $\tau_N$ varying between $0.02-0.1\ \text{sec}$ \cite{Sabti:2020yrt, Boyarsky:2020dzc}; see Appendix \ref{app:decay}.

\section{Can dark decay relax the BBN bound?}

Now let us consider the possibility that HNLs may also decay via additional nonstandard interactions to some dark sector particles.
For fixed $U_{\alpha}$ and $m_N$, a dark decay mode of $N$ would enlarge its total decay width, resulting in a shorter lifetime.
Therefore, one may naively expect that adding new dark sector decay modes can evade or relax the BBN constraint (see e.g., the discussion in Ref.~\cite{Abdullahi:2022jlv} and references therein).
The general reasoning is that decays into truly dark states avoid direct electromagnetic/charged injections and can reduce visible damage to light-element abundances or CMB signals; see e.g., Refs.~\cite{Hufnagel:2017dgo, Hufnagel:2018bjp, Berlin:2019pbq, Depta:2020zbh, Hambye:2021moy, Alonso-Alvarez:2022uxp, Abazajian:2023reo}.
However, in what follows, we show that this, in general, is not true for HNL decays, especially in the region of interest to laboratory HNL searches.

Let us first consider a relativistic dark decay mode, $N\to X$, where $X$ denotes dark sector final states which we assume to consist of relativistic species only.
Hence, once produced, $X$ would behave as dark radiation.
We denote the total decay width of $N$ by $\Gamma_{N}\equiv 1/\tau_N$, the SM contribution to it by $\Gamma_{N}^{({\rm SM})}$, and the decay width of $N\to X$ by $\Gamma_{N\to X}$.
The dark branching ratio thus reads
\begin{equation}
    \text{Br}_{X}\equiv\frac{\Gamma_{N\to X}}{\Gamma_{N}}\thinspace.
    \label{eq:BrX}
\end{equation}

After $N$ has decoupled from the thermal bath [see Eq.~\eqref{eq:Tfo}], it starts free-streaming relativistically until its momentum is red-shifted to a low value comparable to $m_{N}$.
Due to the smallness of the mixing, $N$ decays non-relativistically in most cases.
From relativistic free-streaming to non-relativistic decay, the energy density of $N$ is approximately given by (see Appendix \ref{app:analytical} for derivation):
\begin{equation}
    \rho_{N}\approx\begin{cases}
        \frac{7\pi^{2}}{120}\left(\frac{a_{i}}{a}\right)^{4}T_{i}^{4} & (a\lesssim3.15\frac{a_{i}T_{i}}{m_{N}})\\[2mm]
        \frac{3\zeta(3)}{2\pi^{2}}\left(\frac{a_{i}}{a}\right)^{3}m_{N}T_{i}^{3}e^{-\Gamma_{N}t} & (a\gtrsim3.15\frac{a_{i}T_{i}}{m_{N}})
    \end{cases}\thinspace.
    \label{eq:rho-N}
\end{equation}
Here $a$ is the scale factor and $\zeta(3)\approx 1.202$ is the Riemann zeta function.

\begin{figure}[t!]
    \centering
    \includegraphics[width=\columnwidth]{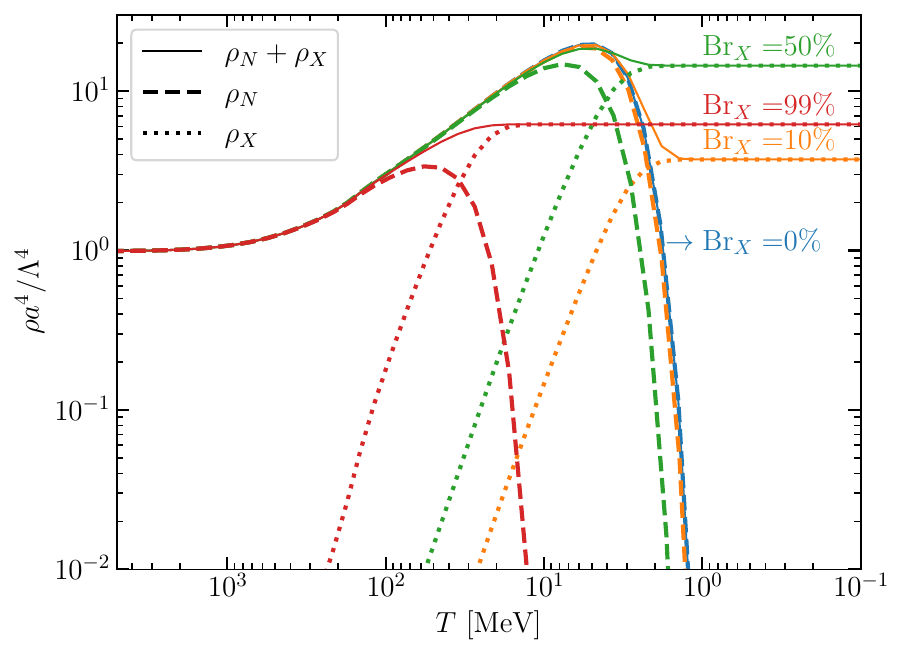}
    \caption{The evolution of energy density of $N$ (dashed lines), its decay product $X$ (dotted lines), and the total energy density of $N$ and $X$ (solid lines).
    The $y$-axis has been normalized to a dimensionless form using $\Lambda^{4}\equiv\rho_{N,i}a_{i}^{4}$.
    For shown examples, we take $m_{N}=1.08$ GeV, $U_{e}^{2}=10^{-10}$, and $\text{Br}_{X}\in\{0\%, 10\%, 50\%, 99\%\}$.}
    \label{fig:rho}
\end{figure}

The first case in Eq.~\eqref{eq:rho-N} is applicable to relativistic free-streaming, during which the comoving energy density $\rho_{N}a^{4}$ remains constant.
When $N$ becomes non-relativistic, corresponding to the second case in Eq.~\eqref{eq:rho-N}, $\rho_{N}$ scales as $a^{-3}$ if $\Gamma_{N}t\ll1$, implying that $\rho_{N}a^{4}$ increases during this period.
At $t\sim1/\Gamma_{N}$, the exponential decay factor $e^{-\Gamma_{N}t}$ in Eq.~\eqref{eq:rho-N} starts to suppress the energy density.
The dashed curves in Fig.~\ref{fig:rho} show the exact evolution of $\rho_{N}a^{4}$ obtained by numerically solving the Boltzmann equation of $N$.
They agree well with our expectation of the behavior of $\rho_{N}$ described above.

The decay of $N$ leads to the production of $X$, whose energy density can also be calculated analytically (see Appendix \ref{app:analytical}).
After a sufficiently long period of evolution, $\rho_{X}$ scales as $a^{-4}$, corresponding to the flat part of the dotted lines in Fig.~\ref{fig:rho}.

Although introducing a dark decay mode makes $N$ decay faster, the total energy stored in the BSM sector at the BBN epoch is not necessarily reduced.
This is illustrated by the solid lines in Fig.~\ref{fig:rho}.
For instance, when $\text{Br}_{X}$ increases from $10\%$ (blue solid line) to $50\%$ (orange solid line), the total energy at $T\lesssim1\ \text{MeV}$ increases significantly.
This significant amount of dark radiation would modify the Hubble expansion rate and thereby alter BBN predictions, as we discuss below.

It is noteworthy that for the shown examples, the final value of $(\rho_{N}+\rho_{X})a^{4}$ is higher than its initial.
This has an important implication: the energy in the dark sector obtained in this way is even higher than what can be carried by a hypothetical neutrino species decoupled at $T_{i}$.
If there was no variation of $g_{\star}$ during the evolution, these examples would lead to $\Delta N_{{\rm eff}}>1$, which would be excluded by BBN and CMB observations \cite{Planck:2018vyg}.
Including the variation of $g_{\star}$ leads to a smaller $\Delta N_{{\rm eff}}$, but still often large enough to be ruled out.
\emph{Therefore, introducing the dark decay mode $N\to X$ with sizable $\text{Br}_{X}$ would generally cause more deviations of BBN predictions from their standard values.}

For $N$ decaying non-relativistically to SM particles and $X$, we find that the resulting $\Delta N_{{\rm eff}}$ can be approximated by (see Appendix \ref{app:analytical})
\begin{equation}
    \Delta N_{{\rm eff}}\approx\frac{\text{Br}_{X}\sqrt{1-\text{Br}_{X}}}{0.16(1-\text{Br}_{X})^{3/2}+0.08g_{\star i}^{13/12}\xi_{N}}\thinspace,
    \label{eq:Neff}
\end{equation}
with $\xi_{N}\equiv\sqrt{\Gamma_{N}^{({\rm SM})}m_{{\rm pl}}}/m_{N}$.
Equation \eqref{eq:Neff} implies that the dependence of $\Delta N_{{\rm eff}}$ on $\text{Br}_{X}$ is not monotonic.
Indeed, as shown in Fig.~\ref{fig:rho}, if one increases $\text{Br}_{X}$ from $50\%$ to $99\%$, the final value of $\rho_{X}a^{4}$ decreases.
One may wonder what would happen if $\text{Br}_{X}$ is increased to a value very close to $100\%$.
In the limit of $\text{Br}_{X}\to1$ with fixed $U_{\alpha}$ and $m_{N}$, Eq.~\eqref{eq:Neff} loses validity because it is derived based on the assumption of non-relativistic decay.
The limit of $\text{Br}_{X}\to1$ implies $\Gamma_{N\to X}\gg\Gamma_{N}^{({\rm SM})}$, drastically enhancing the total decay width $\Gamma_{N}$.
The enhanced $\Gamma_{N}$ causes $N$ to decay before it evolves into the non-relativistic regime, eliminating the rising part of the red solid curve in Fig.~\ref{fig:rho}.
Consequently, the curve of $(\rho_{N}+\rho_{X})a^{4}$ becomes almost flat.
The relativistic decay of $N$ with an almost $100\%$ branching ratio to $X$ simply converts $\rho_{N}$ to $\rho_{X}$ without gain or loss.

Before ending this section, we would like to discuss two possible variants:
\begin{itemize}
    \item $X$ being a massive but stable species;
    \item $X$ being an unstable species.
\end{itemize}
Compared to the case of $X$ being dark radiation, the first variant (see e.g., Refs.~\cite{Barman:2022scg, Li:2022bpp}) would result in more dark-sector energy at the BBN epoch.
This is due to the dilution-resistant effect of the mass\,---\,the energy stored in the form of mass cannot be diluted by the cosmological expansion, as opposed to that stored in the form of radiation\,---\,see Ref.~\cite{Li:2023puz} for a detailed discussion.
So in this case, the solid lines in Fig.~\ref{fig:rho} would exhibit a second rise caused by the mass of $X$, implying that the scenario could be more easily excluded by BBN observations, though in this case its effect cannot be fully accounted for by $\Delta N_{{\rm eff}}$.

If $X$ is unstable, the result crucially depends on whether it dominantly decays to SM particles or other dark species.
For the latter, the conclusion is similar: the dark-sector energy at the BBN epoch cannot be effectively reduced, rendering the BBN constraints more restrictive.
If it is the former (as in the case of HNL decays to explain the short-baseline anomalies \cite{Bertuzzo:2018itn, Arguelles:2018mtc, Abdallah:2024uby}), one can indeed alleviate the BBN constraint, provided that most of the energy in the dark sector is released before the BBN epoch.
But this essentially corresponds to opening new channels for $N$ to release its energy into the SM sector.
We will briefly comment on it later.

\section{BBN and CMB observables}\label{sec:observables}

The effect of HNLs and their decay products on cosmological observables is twofold:
(A) they contribute to the total energy density and thereby affect cosmological expansion;
(B) they may directly participate in some particle physics processes that can affect cosmological observables.

For effect (A), the new physics contribution could be quantified by $N_{{\rm eff}}$ if the new particles stay relativistic during all relevant epochs from neutrino decoupling to recombination.
For decaying non-relativistic HNLs, this requirement is often not satisfied.
However, within the parameter space considered in this work, HNLs must have exhaustively decayed before recombination.
Assuming $X$ is dark radiation, the impact on CMB observations can be fully accounted for by the $N_{{\rm eff}}$ used in CMB data analyses.
To avoid potential confusion, we denote this $N_{{\rm eff}}$ by $N_{{\rm eff}}^{\text{CMB}}$.
From Planck+BAO data \cite{Planck:2018vyg}: 
\begin{equation}
    N_{{\rm eff}}^{\text{CMB}}=2.99\pm0.17\thinspace.
    \label{eq:Neff-exp}
\end{equation}

\begin{figure}[t!]
    \centering
    \includegraphics[width=\columnwidth]{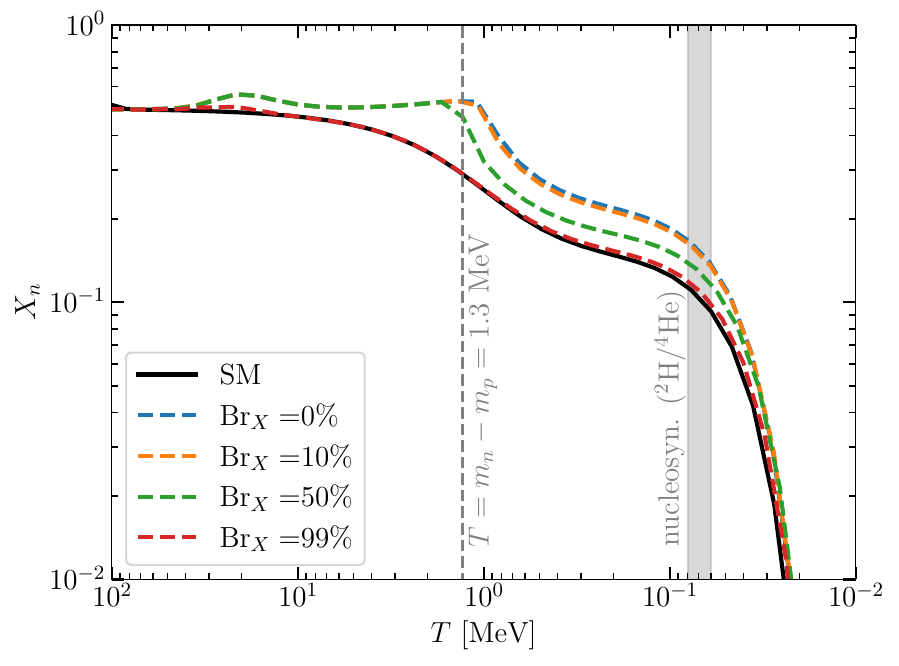}
    \caption{The fractional neutron abundance $X_{n}\equiv n_{n}/(n_{n}+n_{p})$ in the standard (solid line) and HNL decay scenarios (dashed lines).
    For the latter, we assume $m_{N}=1.08$ GeV, $U_{e}^{2}=10^{-10}$, and $\text{Br}_{X}\in\{0\%, 10\%,50\%,99\%\}$.}
    \label{fig:Xn}
\end{figure}

During the BBN epoch, altered cosmological expansion also influences the evolution of the fractional neutron abundance $X_{n}\equiv n_{n}/(n_{n}+n_{p})$, where $n_{p}$ and $n_{n}$ denote the proton and neutron number densities, respectively.
Typically, the freeze-out of neutron-proton conversion processes ($n+\nu_{e}\leftrightarrow p+e^{-}$, $n+e^{+}\leftrightarrow p+\overline{\nu}_{e}$) is advanced due to the faster expansion caused by the new physics contribution to the total energy, giving rise to a higher freeze-out value of $X_{n}$; see Fig.~\ref{fig:Xn}.
Consequently, the helium abundance $Y_{P}$, which is approximately proportional to the freeze-out value of $X_{n}$, increases.

For effect (B), the most prominent example is meson-driven $n$-$p$ conversion.
Among the decay products of HNLs, mesons including pions and kaons have the largest cross sections with baryons.
Therefore, processes like $\pi^{-}+p\to n+\pi^{0}/\gamma$ and $\pi^{+}+n\to p+\pi^{0}$ can efficiently maintain a balance between the neutron and proton abundance, tending to pull $X_{n}$ towards $1/2$.

To quantitatively evaluate the new physics effects, we solve the Boltzmann equations of $\rho_{N}$, $\rho_{X}$, and $X_{n}$ numerically.
They are formulated as follows:
\begin{align}
    \frac{d\rho_{N}}{dt} & =-3H(1+w)\rho_{N}-\rho_{N}\Gamma_{N-}+\rho_{N}^{(\text{eq})}\Gamma_{N+}\thinspace, \label{eq:Boltzmann-N}\\
    \frac{d\rho_{X}}{dt} & =-4H\rho_{X}+\rho_{N}\Gamma_{N-}\text{Br}_{X}\thinspace, \label{eq:Boltzmann-X}\\
    \frac{dX_{n}}{dt} & =-\Gamma_{n\to p}X_{n}+\Gamma_{p\to n}\left(1-X_{n}\right)\thinspace, \label{eq:Boltzmann-Xn}
\end{align}
where $w$ is the equation of state of $N$, $\Gamma_{N\pm}$ are production and depletion rates of $N$, $\rho_{N}^{(\text{eq})}$ is the thermal equilibrium value of $\rho_{N}$, and $\Gamma_{n\to p(p\to n)}$ are neutron-to-proton (proton-to-neutron) conversion rates.
The SM contributions to $\Gamma_{n\to p(p\to n)}$ can be found in Refs.~\cite{Baumann:2022mni,Meador-Woodruff:2024due}.
Additionally, there are also meson-driven contributions from pions and kaons generated by HNL decays.
We detail the calculation of these contributions in Appendix \ref{app:meson-driven}.
When solving the Boltzmann equations, we replace $dt\to da/(aH)$ with $H$ including both SM and BSM contributions.
Here we do not include backreaction terms responsible for energy transfer from $X$ to $N$.
The effect of such terms is expected to be insignificant as long as the temperature of the dark sector is low, which is guaranteed by $N_{\rm eff}$ constraints.
By solving Eqs.~\eqref{eq:Boltzmann-N}-\eqref{eq:Boltzmann-Xn}, we obtain the evolutions of $\rho_{N,X}$ and $X_{n}$ shown in Figs.~\ref{fig:rho} and \ref{fig:Xn}, respectively.

\begin{figure}[t]
    \centering
    \hspace{-5ex}\includegraphics[width=0.5\textwidth]{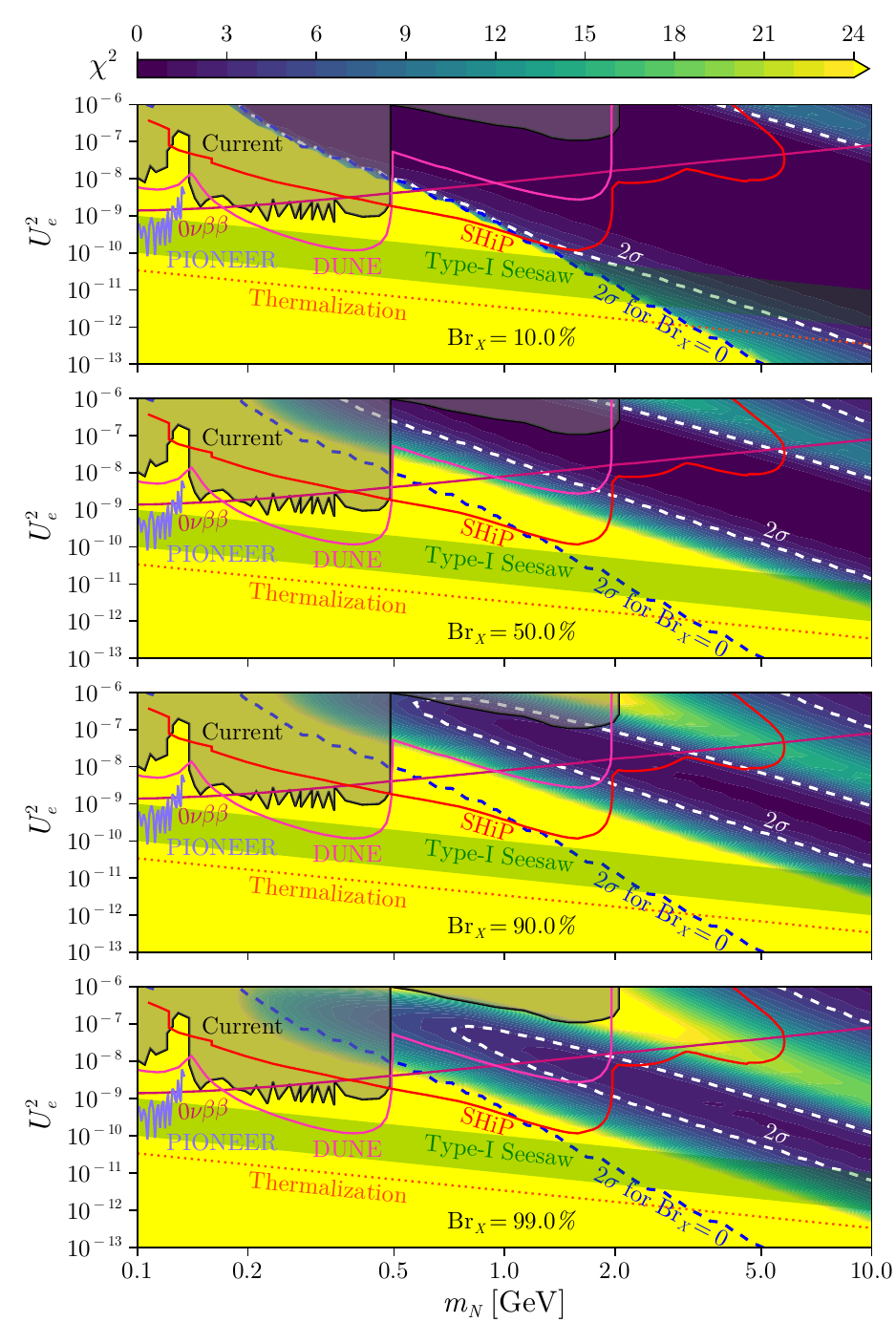}
    \caption{%
        The combined $\chi^{2}$ values to reveal the cosmologically allowed (blue) and disfavored (yellow) parameter space for different ${\rm Br}_X$ values.
        The current laboratory constraints and future sensitivities, as well as the seesaw band and thermalization line, are shown for comparison.
    }
    \label{fig:chi2}
\end{figure}

In the SM, when neutrons and protons are in chemical equilibrium, the neutron-to-proton ratio is proportional to $e^{-Q/T}$ where $Q=m_{n}-m_{p}\approx1.3$ MeV.
The chemical equilibrium is no longer maintained when the exponential suppression $e^{-Q/T}$ becomes significant.
Hence after $T\lesssim Q$, $X_{n}$ approximately approaches a constant until neutron decay becomes significant.
Since the neutron lifetime is $\tau_{n}\approx877\ \text{sec}$ \cite{Fuwa:2024cdf}, most neutrons decay roughly at $t\sim\tau_{n}$, corresponding to $T\sim0.03$ MeV according to Eq.~\eqref{eq:t-T}.
Shortly before the decay, nucleosynthesis produces deuterium ($^{2}\text{H}$) via $n$-$p$ fusion and then helium ($^{4}\text{He}$) from deuterium.
This happens at $t\sim 200-300\ \text{sec}$, represented by the vertical gray band in Fig.~\ref{fig:Xn} with $T\in [0.06,\, 0.08]$ MeV.

The relic abundance of helium after nucleosynthesis is approximately determined by Ref.~\cite{Dodelson:2020bqr}
\begin{equation}
    Y_{P} \approx 2X_{n}(t_{{\rm nuc}})\thinspace,
    \label{eq:Yp}
\end{equation}
where $t_{{\rm nuc}}\approx250\ \text{sec}$ is the typical nucleosynthesis time.

In the presence of HNLs, the meson-driven processes pull $X_{n}$ towards a value slightly higher than $1/2$, as can be seen from the colored dashed lines in Fig.~\ref{fig:Xn}.
This effect persists until all HNLs and the produced mesons have decayed.
The blue/orange/green lines, corresponding to $\text{Br}_{X}=0\%/10\%/50\%$, have sufficiently long-lived HNLs such that the pulling effect lasts until $T\lesssim Q$.
Consequently, their $X_{n}$ at $t_{{\rm nuc}}$ is significantly enhanced compared to the SM value (black line).
Increasing $\text{Br}_{X}$ to $99\%$, which implies a much shorter HNL lifetime, results in the red line, which exhibits a small deviation from the SM curve at $T\sim20$ MeV.
However, this deviation is rapidly washed out in the subsequent evolution by SM thermal processes.
Nevertheless, this does not necessarily imply that all new physics effects in this case are eliminated.
If the decay width of $N\to X$ is large, the background receives considerable corrections from the dark sector\,---\,see effect (A) above.
Therefore, at $T\lesssim Q$, the red curve starts to deviate from the SM curve again.
This deviation, though small (about $6.5\%$), is still sufficient to be observed, since current observations have measured the primordial helium abundance to a percent level \cite{Izotov:2014fga, Aver:2015iza, Valerdi:2019beb, ParticleDataGroup:2024cfk}:
\begin{equation}
    Y_{P}=0.245\pm0.003\thinspace.
    \label{eq:Yp-exp}
\end{equation}
We have used the PDG recommended value instead of the latest measurement of $Y_P$ with higher precision \cite{Yanagisawa:2025mgx} (which could have strengthened our bounds) because the latter has mild tension with the SM and Planck results.

To quantitatively evaluate how much the resulting $\Delta N_{{\rm eff}}^{\text{CMB}}$ and $Y_{P}$ are disfavored by cosmological observations, we employ the following $\chi^2$ function:
\begin{equation}
    \chi^{2}=\left(\frac{\Delta N_{{\rm eff}}^{\text{CMB}}}{\sigma_{N}}\right)^{2}+\left(\frac{\Delta Y_{P}}{\sigma_{Y}}\right)^{2},
    \label{eq:chi2}
\end{equation}
where $\Delta N_{{\rm eff}}^{\text{CMB}}$ and $\Delta Y_{P}$ denote the new physics contributions to $N_{{\rm eff}}^{\text{CMB}}$ and $Y_{P}$, respectively.
Their uncertainties, denoted by $\sigma_{N}$ and $\sigma_{Y}$, are indicated by Eqs.~\eqref{eq:Neff-exp} and \eqref{eq:Yp-exp}, respectively.
We perform a numerical scan of the parameter space with $m_{N}\in[0.1,\ 10]\ \text{GeV}$
and $U_{e}^{2}\in[10^{-13},\ 10^{-6}]$ (similar results can be obtained for $U_\mu^2$ and $U_\tau^2$), and present the result in Fig.~\ref{fig:chi2}.
In this plot, the yellow region is strongly disfavored by cosmological observation at a confidence level (C.L.) above $5 \sigma$ ($\chi^{2}\geq25$).
A less conservative constraint is represented by the white dashed contour, corresponding to $2\sigma$ C.L.
For comparison, we also add the standard $2\sigma$ BBN bound for $\text{Br}_{X}=0$ on each panel, presented as the blue dashed line, which is always found to be less restrictive than the white dashed contour.
For very small mixing, HNLs may never have reached thermal equilibrium.
In Fig.~\ref{fig:chi2}, we add a thermalization line below which HNLs do not thermalize (see Appendix \ref{app:freeze-in} for an analytical derivation).
The part below this line corresponds to the freeze-in regime.

Also shown in Fig.~\ref{fig:chi2} is the Type-I seesaw band, plotted using Eq.~\eqref{eq:seesaw} with $m_{\nu}$ varying from $0.01$ eV to $0.1$ eV.
Targeting this theoretical benchmark, current laboratory bounds and future sensitivity reaches are also shown.
The current laboratory bounds are taken from Refs.~\cite{Bolton:2019pcu, website1, website2} presented as gray shaded regions.
We also show a bound from $0\nu\beta\beta$ experiments, which is however not shaded because it can be significantly weakened in the presence of more than one HNLs \cite{Bolton:2019pcu}.
For future experiments, we select three representative ones, namely, SHiP \cite{Alekhin:2015byh, SHiP:2018xqw}, DUNE \cite{Krasnov:2019kdc, Ballett:2019bgd, Berryman:2019dme}, and PIONEER \cite{PIONEER:2022yag}, as shown by the solid lines, which could potentially reach the seesaw band.
However, we find that the laboratory-accessible parts of the seesaw band are strongly disfavored by cosmological data.
Our results suggest that the possible loophole of introducing dark decay channels to evade the cosmological bounds on HNLs is firmly ruled out.
Instead, dark decay modes strengthen the cosmological constraints.

\section{Discussion and summary}

We now comment on some possible scenarios that might evade the cosmological bounds on HNLs:
(i) The most straightforward scenario that one can imagine is that $N$ and $X$ can rapidly release their energy back into the SM thermal bath before the BBN epoch.
However, this scenario would imply additional interactions of the dark sector with the SM.
And the interaction strength cannot be too small for it to play a significant role.
This might cause $N$ to interact more with the SM via the dark sector than via the mixing with active neutrinos.
(ii) Another possible scenario is that $X$ contains both SM and dark-sector particles.
For instance, in Ref.~\cite{Deppisch:2024izn}, the dominant HNL decay channel is $N\to \nu a$ where $a$ is an axion-like particle.
This channel efficiently depletes the energy of $N$ when it becomes non-relativistic, leaving only the SM thermal plasma and $a$ in the BBN epoch, with the number density of $a$ lower than the neutrino one.
(iii) In addition to modifying HNL decay, it is also possible to relax the BBN constraint via entropy dilution \cite{Fuller:2011qy}.
As one can see from Eq.~\eqref{eq:Neff}, if the thermal bath contains a very large number of species when $N$ decouples (i.e., very large $g_{\star,i}$), its effect at the BBN epoch would be significantly diluted.
(iv) HNLs with very small mixing may have never been in thermal equilibrium and can also evade the BBN bound.
This is the freeze-in regime (see Ref.~\cite{Chen:2024cla} for a recent study), in contrast to the freeze-out regime considered here.
Note however that the photodisintegration effect on formed nuclei in this regime can be significant and may set strong constraints \cite{Domcke:2020ety}.
The very small mixing renders HNLs almost stable, allowing keV-scale HNLs to serve as dark matter \cite{Dodelson:1993je, Shi:1998km, DeGouvea:2019wpf, Astros:2023xhe, Dev:2025sah}.
Besides, the mixing could be suppressed by new interactions contributing to the thermal effective potential of HNLs, which has been employed to evade the cosmological bounds on eV-scale HNLs \cite{Hannestad:2013ana, Dasgupta:2013zpn, Saviano:2014esa, Chu:2015ipa, Forastieri:2017oma, Chu:2018gxk}.
(v) Another way to avoid the BBN bound is to have a very low reheating temperature such that the HNL contribution to the radiation energy density could be reduced \cite{Gelmini:2004ah, Gelmini:2008fq, Gelmini:2019wfp, Hasegawa:2020ctq}.
However, the impact of low reheating temperature on cosmological observables has to be kept in mind \cite{Barbieri:2025moq}.

In conclusion, we emphasize that modern precision cosmology imposes strong bounds on HNLs in the parameter space relevant to laboratory searches.
These bounds cannot be alleviated simply by introducing new dark decay channels.
If future laboratory searches detect a positive HNL signal in the cosmologically disfavored regime, it would imply some nonstandard cosmology or a nontrivial HNL scenario.

\acknowledgments

BD thanks the IHEP Theory Group for warm hospitality where this work was initiated.
BD also wishes to acknowledge the Center for Theoretical Underground Physics and Related Areas (CETUP*) and the Institute for Underground Science at SURF, as well as the Galileo Galilei Institute for Theoretical Physics (GGI), for hospitality and for providing a stimulating environment during the completion of this work.
The work of BD was partly supported by the U.S. Department of Energy under grant No.~DE-SC0017987, and by a Humboldt Fellowship from the Alexander von Humboldt Foundation.
XJX is supported in part by the National Natural Science Foundation of China (NSFC) under grant No.~12141501 and also by the CAS Project for Young Scientists in Basic Research (YSBR-099).

\section*{Data Availability}

The data that support the findings of this article are openly available \cite{Wu:2026Data}.

\appendix

\section{HNL Decay to SM Particles}\label{app:decay}

\begin{figure*}
    \centering
    \includegraphics[width=\textwidth]{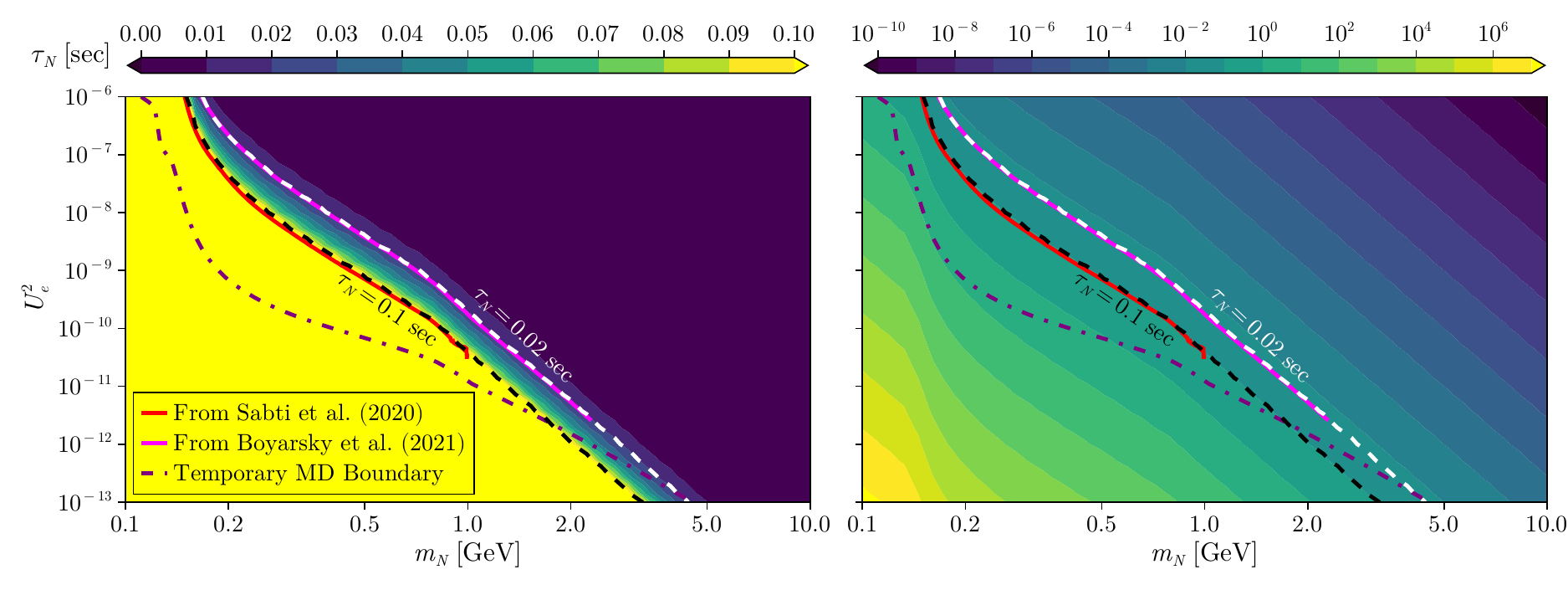}
    \caption{The lifetime $\tau_{N}$ as a function of the mixing $U_{e}$, and the mass, $m_{N}$, assuming the HNL only mixes with $\nu_{e}$.
    The left panel presents $\tau_N$ in linear scales, and the right panel in log scales.
    The white and black dashed lines correspond to $\tau_{N}=0.02\ \text{sec}$ and $0.1\ \text{sec}$, respectively.
    The purple dash-dotted line represents the boundary of temporary matter domination caused by HNLs, obtained using Eq.~\eqref{eq:md-U}.
    The cosmological bounds from Sabti et al.~(2020)\cite{Sabti:2020yrt} and Boyarsky et al.~(2021)\cite{Boyarsky:2020dzc} can be well approximated by the two $\tau_{N}$ contours.}
    \label{fig:lifetime}
\end{figure*}

For $m_N<m_W$, HNLs can decay into various leptonic and hadronic modes, among which the most relevant ones are $N\to3\nu$, $N\to\nu+\ell^{\pm}+\ell^{\mp}$, $N\to\ell^{\mp}+\mathfrak{m}^{\pm}$, and $N \to\nu+\mathfrak{m}^{0}$, where $\mathfrak{m}^{\pm,0}$ denotes charged and neutral mesons (e.g., $\pi^{\pm}$ and $\pi^{0}$).
The invisible decay mode $N\to3\nu$ has a compact and simple expression for its decay width\cite{Bondarenko:2018ptm}:
\begin{align}
    \Gamma_{N\to3\nu} & =\frac{G_{F}^{2}m_{N}^{5}}{192\pi^{3}}\sum_{\alpha}U_{\alpha}^{2}\thinspace,
    \label{eq:-25}
\end{align}
where $G_{F}$ is the Fermi constant.
Other decay channels are more complicated due to the involved mass thresholds and hadronic matrix elements.
We refer to Ref.~\cite{Bondarenko:2018ptm} for a comprehensive calculation of all HNL decay modes.
Here we adopt a simple way to calculate the total HNL decay width for SM final states:
\begin{equation}
    \Gamma^{\rm (SM)}_{N}=\Gamma_{N\to3\nu}/\text{Br}_{N\to3\nu}\thinspace,
    \label{eq:Br-nu}
\end{equation}
where $\text{Br}_{N\to3\nu}$ is the branching ratio of the invisible mode presented in Fig.~18 of Ref.~\cite{Bondarenko:2018ptm}.
Figure~\ref{fig:lifetime} shows the lifetime $\tau_{N}=1/\Gamma_{N}^{\rm (SM)}$ obtained using Eqs.~\eqref{eq:-25} and \eqref{eq:Br-nu}, as a function of $U_{\alpha}$ and $m_{N}$.
We present $\tau_{N}$ in both linear (left panel) and log scales (right panel) because the former is more relevant to the BBN epoch and the latter offers a broader scope, allowing us to relate the lifetime with other potentially relevant epochs in the Universe.
The black dot-dashed and the white dashed lines correspond to $\tau_{N}=0.1\ \text{sec}$ and $0.02\ \text{sec}$, respectively.
Imposing the BBN bounds obtained in Refs.~\cite{Sabti:2020yrt, Boyarsky:2020dzc} on this plot, we can see that they can be well approximated by the lifetime contours of $\tau_{N}=0.1\ \text{sec}$ and $0.02\ \text{sec}$.

\section{Analytical estimate in the freeze-out regime}\label{app:analytical}

In the freeze-out regime, HNLs decouple from the thermal bath at a certain freeze-out temperature.
Below this temperature, HNLs start to free-stream from relativistic states to non-relativistic states, and eventually decay to the SM particles and the dark sector final states $X$, if any.

For decoupled HNLs, when the decay is insignificant, the number density is given by
\begin{equation}
    n_{N}=n_{N,i}\left(\frac{a_{i}}{a}\right)^{3},
    \label{eq:A}
\end{equation}
where the subscript $i$ denotes an initial point which can be set at decoupling.
The energy density is approximately given by
\begin{equation}
    \rho_{N}\approx\begin{cases}
        2\times\frac{7}{8}\times\frac{\pi^{2}}{30}T_{N}^{4} & (\text{UR})\\
        m_{N}n_{N} & (\text{NR})
    \end{cases}\thinspace.
    \label{eq:A-1}
\end{equation}
Here UR and NR indicate ultra-relativistic and non-relativistic limits, respectively.
The temperature of HNLs, $T_{N}$, is determined by
\begin{align}
    T_{N} & =T_{i}\frac{a_{i}}{a}\thinspace,
    \label{eq:A-2}
\end{align}
where $T_{i}$ is the temperature of HNLs at decoupling (also the temperature of the thermal bath).
Note that after decoupling, $T_{N}$ may be lower than the thermal bath temperature $T$, which is affected by the variation of $g_{\star}$:
\begin{equation}
    T=\left(\frac{g_{\star,i}}{g_{\star}}\right)^{\frac{1}{3}}T_{i}\frac{a_{i}}{a}\thinspace.
    \label{eq:A-3}
\end{equation}
Strictly speaking, one should use the effective number of entropy degrees of freedom ($g_{\star s}$) in the above expression.
However, since the difference between $g_{\star s}$ and $g_{\star}$ is small above the MeV scale, we neglect it in the analytical calculation.

To estimate the transition point from the UR to the NR regime, we equate the UR and NR expressions in Eq.~\eqref{eq:A-1} and obtain the following solution:
\begin{equation}
    a_{\text{tr}}\approx\frac{7\pi^{4}}{180\zeta(3)}\frac{a_{i}T_{i}}{m_{N}}\approx3.15\frac{a_{i}T_{i}}{m_{N}}\thinspace,
    \label{eq:A-4}
\end{equation}
where $\zeta$ is the Riemann zeta function and $\zeta(3)\approx1.202$ arises from computing the relativistic fermion number density $n_{N,i}$.
For $a\lesssim a_{\text{tr}}$ or $a\gtrsim a_{\text{tr}}$, $\rho_{N}$ can be approximately calculated using the UR or NR limit, respectively.

In most cases, $N$ decays in the non-relativistic regime, allowing us to take the decay effect into account via
\begin{equation}
    \rho_{N}\to\rho_{N}e^{-\Gamma_{N}t}\thinspace,
    \label{eq:A-5}
\end{equation}
where $\Gamma_{N}$ is the total decay rate of $N$ at rest.
Equation \eqref{eq:A-5} only applies to non-relativistic decay.
For relativistic decay, the energy spectrum of $N$ would be distorted, requiring a much more
complex treatment\,---\,see e.g., Appendix C in Ref.~\cite{Wu:2024uxa}.

Assembling the above pieces together, we obtain the following analytical expression for $\rho_{N}$:
\begin{equation}
    \rho_{N}\approx\begin{cases}
        \frac{7\pi^{2}}{120}\left(\frac{a_{i}}{a}\right)^{4}T_{i}^{4} & (a\lesssim a_{\text{tr}})\\[2mm]
        \frac{3\zeta(3)}{2\pi^{2}}\left(\frac{a_{i}}{a}\right)^{3}T_{i}^{3}e^{-\Gamma_{N}t} & (a\gtrsim a_{\text{tr}})
    \end{cases}\thinspace.
    \label{eq:A-1-1}
\end{equation}

Since $\rho_{N}$ scales as $a^{-3}$ in the non-relativistic
regime before it decays while $\rho_{{\rm SM}}$ roughly scales as
$a^{-4}$, it is possible that at some point $\rho_{N}$ may exceed
$\rho_{{\rm SM}}$, causing a temporary period of matter domination
(MD).
To estimate the condition for MD, we compute the ratio
$\rho_{N}/\rho_{{\rm SM}}$ using Eq.~\eqref{eq:A-1-1} and find its
maximum, which is given by
\begin{equation}
\left(\frac{\rho_{N}}{\rho_{{\rm SM}}}\right)_{{\rm max}}\approx0.43\frac{m_{N}}{g_{\star s,i}\sqrt{\Gamma_{N}m_{{\rm pl}}}}\left(\frac{g_{\star,s}}{g_{\star}^{3/4}}\right)_{{\rm decay}}\thinspace,\label{eq:rho-N-SM}
\end{equation}
where the factor $\left(g_{\star,s}/g_{\star}^{3/4}\right)_{{\rm decay}}$
should be evaluated at $t\sim1/\Gamma_{N}$.

Substituting Eq.~\eqref{eq:Br-nu} into Eq.~\eqref{eq:rho-N-SM} and solving $\left(\rho_{N}/\rho_{{\rm SM}}\right)_{{\rm max}}=1$ with respect to $U_{\alpha}^{2}$, we obtain
\begin{equation}
U_{\alpha}^{2}\approx6.8\times10^{-7}\text{Br}_{N\to3\nu}\left(\frac{\text{GeV}}{m_{N}}\right)^{3}\frac{1}{g_{\star s,i}^{2}}\left(\frac{g_{\star,s}}{g_{\star}^{3/4}}\right)_{{\rm decay}}^{2}\thinspace.\label{eq:md-U}
\end{equation}
This is the critical value of $U_{\alpha}^{2}$ to reach a temporary
MD period.
In Fig.~\ref{fig:lifetime}, we plot it as a purple dash-dotted line.
Below this line, the universe undergoes a short period of MD
before HNLs decay.

The energy density of the dark radiation $X$ is governed by the following Boltzmann equation,
\begin{equation}
    \frac{d\rho_{X}}{dt}+4H\rho_{X}=\rho_{N}\Gamma_{N}\text{Br}_{X}\thinspace,
    \label{eq:A-6}
\end{equation}
where $\text{Br}_{X}$ is the branching ratio of the dark decay mode.
Since the left-hand side can be written as $\frac{d(\rho_{X}a^{4})}{a^{4}dt}$ and $dt=H^{-1}d\ln a$, Eq.~\eqref{eq:A-6} implies
\begin{equation}
    \rho_{X}a^{4}=\Gamma_{N}\text{Br}_{X}\int\frac{a^{4}}{H}\rho_{N}d\ln a\thinspace.
    \label{eq:A-7}
\end{equation}
Using the NR expression in Eq.~\eqref{eq:A-1-1} and $H\approx H_{i}a_{i}^{2}/a^{2}$, one can calculate the above integral analytically.
The result reads
\begin{equation}
    \rho_{X}\approx\frac{3\zeta(3)}{4\pi^{2}}\left(\frac{a_{i}}{a}\right)^{4}\sqrt{\frac{\pi}{\gamma}}\text{Br}_{X}m_{N}T_{i}^{3}{\cal F}\left(\frac{a}{a_{i}}\sqrt{\gamma}\right),
    \label{eq:A-8}
\end{equation}
with
\begin{equation}
    \gamma\equiv\frac{\Gamma_{N}}{2H_{i}}\thinspace.
    \label{eq:A-12}
\end{equation}
The ${\cal F}$ function is defined as
\begin{equation}
    {\cal F}(x)\equiv\text{erf}(x)-\frac{2}{\sqrt{\pi}}xe^{-x^{2}}\thinspace,
    \label{eq:A-9}
\end{equation}
which varies from $0$ to $1$ for $x\in[0,\ \infty)$.

The above calculation can be straightforwardly adapted for computing the energy injection into the SM thermal plasma.
Since the branching ratio of $N\to{\rm SM}$ is simply $1-\text{Br}_{X}$, the SM energy density
receives the following contribution:
\begin{equation}
    \Delta\rho_{\text{SM}}\approx\frac{3\zeta(3)}{4\pi^{2}}\left(\frac{a_{i}}{a}\right)^{4}\sqrt{\frac{\pi}{\gamma}}\left(1-\text{Br}_{X}\right)m_{N}T_{i}^{3}{\cal F}\left(\frac{a}{a_{i}}\sqrt{\gamma}\right).
    \label{eq:A-10}
\end{equation}
In particular, the neutrino energy density of a single flavor is given by
\begin{equation}
    \rho_{\nu} =\frac{7\pi^{2}}{120}\left(\frac{g_{\star i}}{g_{\star }}\right)^{\frac{1}{3}}T_{i}^{4}\left(\frac{a_{i}}{a}\right)^{4}+\frac{2\times\frac{7}{8}}{g_{\star}}\Delta\rho_{\text{SM}}\thinspace.
    \label{eq:A-11}
\end{equation}
Equation \eqref{eq:A-11} is valid only before neutrino decoupling.
After neutrino decoupling, $\rho_{\nu}$ scales as $a^{-4}$, provided that the decay has effectively stopped.
If after neutrino decoupling there is still a significant amount of HNLs decaying, then one needs to further calculate the fraction of $\Delta\rho_{\text{SM}}$ that is injected into the neutrino sector.
This is complicated for an analytical estimate, but straightforward for numerical implementation.

The ratio $\rho_{X}/\rho_{\nu}$ after neutrino decoupling can be directly interpreted as the new physics contribution to $N_{{\rm eff}}$.
By combining the above analytical results, we also get an estimate of the contribution to $N_{{\rm eff}}$:
\begin{equation}
    \Delta N_{{\rm eff}}\approx\frac{\text{Br}_{X}}{0.16(1-\text{Br}_{X})+0.15g_{\star i}^{4/3}\gamma^{1/2}\frac{T_{i}}{m_{N}}}\thinspace.
    \label{eq:A-13}
\end{equation}
Using $\Gamma_{N}=\Gamma_{N}^{({\rm SM})}/(1-\text{Br}_{X})$, and $\gamma$ from Eq.~\eqref{eq:A-12} with $H_{i}\approx1.66g_{\star}^{1/2}T_{i}^{2}/m_{{\rm pl}}$, it is straightforward to obtain Eq.~\eqref{eq:Neff}.

\section{The freeze-in regime and thermalization}\label{app:freeze-in}

When the mixing is sufficiently small, HNLs cannot thermalize, implying that they are in the freeze-in regime.
While both freeze-out and freeze-in are included in our numerical code that directly solves the relevant Boltzmann equations, here we would like to provide an analytical estimate of the condition for thermalization.

The production rate of HNLs is approximately given by (see, e.g., Ref.~\cite{Li:2022bpp})
\begin{equation}
    \Gamma_{N,\text{prod}}\approx\theta_{{\rm eff}}^{2}\Gamma_{\nu_{L},{\rm prod}}\thinspace,\label{eq:p}
\end{equation}
with 
\begin{align}
    \Gamma_{\nu_{L},{\rm prod}} & \approx3c_{1}G_{F}^{2}T^{5}\thinspace,\label{eq:p-1}\\
    \sin^{2}2\theta_{{\rm eff}} & \approx\frac{\sin^{2}2\theta}{\sin^{2}2\theta+\left(\cos2\theta-6TV/m_{N}^{2}\right)^{2}}\thinspace,\label{eq:p-2}\\
    V & \approx-c_{2}G_{F}^{2}T^{5}\,.
\end{align}
Here $\theta_{{\rm eff}}$ is the effective mixing angle of $N$ with SM neutrinos ($\nu_{L}$).
It is different from the in-vacuum mixing angle $\theta\approx U_{\alpha}^{1/2}$ due to the thermal MSW potential $V$ in Eq.~\eqref{eq:p-2}.
The two coefficients $c_{1}$ and $c_{2}$ have a mild temperature dependence; however, for a simple estimate we treat them as constants, taking $c_{1}\approx1.27$ and $c_{2}\approx2.5\times10^{2}$.
The production rate is to be compared with $H$.
Hence we define
\begin{align}
    \gamma_{N,\text{prod}} & \equiv \frac{\Gamma_{N,\text{prod}}}{H} \nonumber \\
    & \approx\frac{3c_{1}G_{F}^{2}T^{3}m_{{\rm pl}}\sin^{2}2\theta}{4g_{H}\left(\left(6c_{2}G_{F}^{2}m_{N}^{-2}T^{6}+\cos2\theta\right)^{2}+\sin^{2}2\theta\right)}\thinspace,\label{eq:p-3}
\end{align}
where $g_{H}\equiv m_{{\rm pl}}H/T^{2}\approx1.66g_{\star}^{1/2}$.
The condition for thermalization corresponds to $\gamma_{N,\text{prod}}\gtrsim1$.
Note that $\gamma_{N,\text{prod}}$ asymptotically approaches $T^{-9}$ and $T^{3}$ at very high and low $T$, respectively, implying that it peaks at an intermediate temperature.
Taking $\cos2\theta\to1$ and $\sin2\theta\to0$ in the denominator, the peak temperature and the corresponding peak value of $\gamma_{N,\text{prod}}$ can be calculated analytically:
\begin{align}
    T^{(\text{peak})} & \approx\left(\frac{m_{N}}{3\sqrt{2c_{2}}G_{F}}\right)^{\frac{1}{3}}\approx11\ \text{GeV}\times\left(\frac{m_{N}}{\text{GeV}}\right)^{\frac{1}{3}},\label{eq:p-4}\\
    \gamma_{N,\text{prod}}^{(\text{peak})} & \approx\frac{9c_{1}G_{F}m_{N}m_{{\rm pl}}U_{\alpha}^{2}}{16g_{H}\sqrt{2c_{2}}}\thinspace.\label{eq:p-5}
\end{align}
From Eq.~\eqref{eq:p-5}, one can see that the condition for thermalization
corresponds to
\begin{equation}
    U_{\alpha}^{2}\gtrsim\frac{16g_{H}\sqrt{2c_{2}}}{9c_{1}G_{F}m_{N}m_{{\rm pl}}}\approx3\times10^{-12}\times\left(\frac{g_{\star}}{83.5}\right)^{\frac{1}{2}}\times\frac{\text{GeV}}{m_{N}}\thinspace.\label{eq:p-6}
\end{equation}
Here, $g_{\star}$ is evaluated at the peak temperature.
For $m_{N}=1$
GeV, it is $83.5$, which is the value of $g_{\star}$ at peak temperature of 11 GeV,
as suggested by Eq.~\eqref{eq:p-4}.
For smaller or larger $m_{N}$,
$g_{\star}$ needs to be adjusted accordingly but the variation is
not significant (varying from $82$ to $87$ for $0.1<m_{N}/\text{GeV}<10 $)
within the parameter space considered in this work.

\section{The abundance of HNLs at \texorpdfstring{$T = 1$}{T = 1} MeV}\label{app:abundance}

\begin{figure}[ht]
    \centering
    \includegraphics[width=\columnwidth]{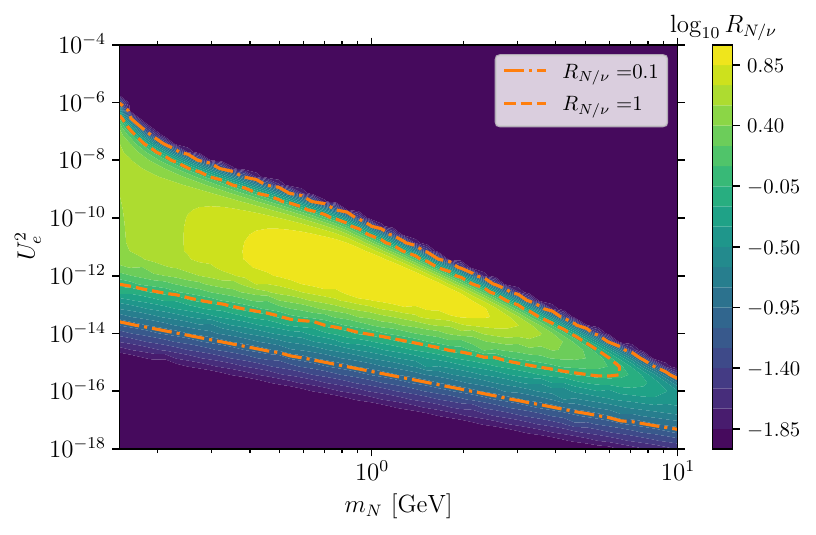}
    \caption{The ratio $R_{N/\nu}\equiv \rho_N/\rho_{\nu}$ at $T = 1$ MeV as an indicator of the abundance of HNLs around neutron freeze-out.}
    \label{fig:energy-ratio}
\end{figure}

The abundance of HNLs at $T = 1$ MeV is useful for understanding their impact on BBN, since this temperature is close to neutron freeze-out.
If the abundance of HNLs at this point is high, the neutron-to-proton ratio is expected to be significantly affected.
In Fig.~\ref{fig:energy-ratio}, we present the ratio $R_{N/\nu}\equiv \rho_N/\rho_{\nu}$ at $T = 1$ MeV, obtained from our numerical calculation assuming ${\rm Br}_X=0$.
This ratio is exponentially suppressed in the large-mixing regime (above the upper branches of the orange curves) due to fast decay.
For very small mixing, it enters the freeze-in regime, in which this ratio is suppressed by the production of HNLs.

\section{Meson-driven \texorpdfstring{$p$-$n$}{p-n} conversion}\label{app:meson-driven}

One of the most significant effects of HNLs on BBN is that mesons produced from $N$ decay may cause meson-driven $p$-$n$ conversion.
Taking $\pi^{\pm}$ for example, the following processes are possible:
\begin{equation}
    \pi^{-}+p\to n+\pi^{0}/\gamma\thinspace,\pi^{+}+n\to p+\pi^{0}\thinspace.
    \label{eq:-28}
\end{equation}
Due to the Sommerfeld enhancement, meson-driven $p\to n$ conversion processes are more efficient than meson-driven $n\to p$ conversion processes.

Note that the lifetime of $\pi^{\pm}$, $\tau_{\pi^{\pm}}\approx2.6\times10^{-8}\ \text{sec}$, is much shorter than the time scale of the early Universe around the BBN epoch.
Due to the short lifetime, if all processes for $\pi^{\pm}$ production have effectively ceased, the abundance of $\pi^{\pm}$ would decrease rapidly via decay.
In the presence of HNLs, which serve as a steady source of $\pi^{\pm}$, the number density $n_{\pi^{\pm}}$ can maintain a balanced value\cite{Boyarsky:2020dzc}:
\begin{equation}
    n_{\pi^{\pm}}\Gamma_{\pi^{\pm}}=n_{N}\Gamma_{N\to\ell^{\mp}+\pi^{\pm}}\thinspace.
    \label{eq:-29}
\end{equation}
The depletion of $\pi^{\pm}$ via Eq.~\eqref{eq:-28} is negligible compared to that via decay.
Equation \eqref{eq:-29} implies that the balanced value of $n_{\pi^{\pm}}$ is proportional to the lifetime of $\pi^{\pm}$.
This further implies that the inverse processes of Eq.~\eqref{eq:-28}, i.e., $n+\pi^{0}\to\pi^{-}+p$ and $p+\pi^{0}\to\pi^{+}+n$, are unimportant because $\pi^{0}$ has a much shorter lifetime ($\tau_{\pi^{0}}\approx8.5\times10^{-17}\ \text{sec}$).

\begin{figure*}
    \centering
    \includegraphics[width=\textwidth]{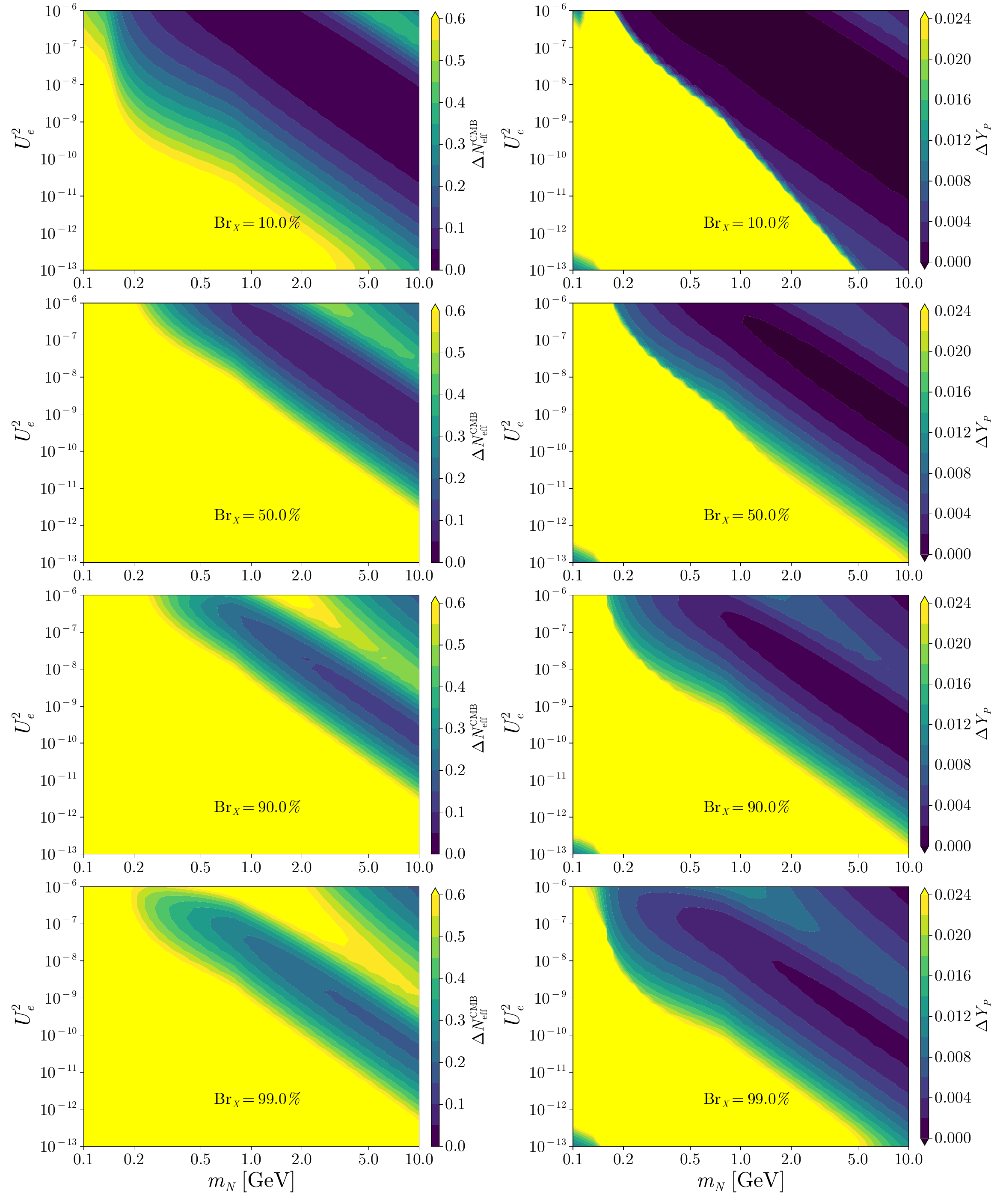}
    \caption{The CMB and BBN observables, $\Delta N_{{\rm eff}}^{{\rm CMB}}$ and $Y_{P}$, affected by HNLs in the presence of a dark-decay channel with the branching ratio $\text{Br}_{X}$.}
    \label{fig:Neff-Yp}
\end{figure*}

The effect of the processes in Eq.~\eqref{eq:-28} on BBN can be taken into account by adding the following contributions to $\Gamma_{n\to p}$ and $\Gamma_{p\to n}$ in the Boltzmann equation of $X_n$ [cf.~Eq.~\eqref{eq:Boltzmann-Xn}]:
\begin{align}
    \Gamma_{n\to p}^{(\pi^{+})} & =n_{\pi^{+}}\left\langle \sigma_{n\to p}^{\pi^{+}}v\right\rangle , \label{eq:np-pi}\\
    \Gamma_{p\to n}^{(\pi^{-})} & =n_{\pi^{-}}\left\langle \sigma_{p\to n}^{\pi^{-}}v\right\rangle , \label{eq:pn-pi}
\end{align}
where the thermally averaged cross sections $\left\langle \sigma_{n\to p}^{\pi^{+}}v\right\rangle$ and $\left\langle \sigma_{p\to n}^{\pi^{-}}v\right\rangle$ can be found in Ref.~\cite{Kohri:2001jx}:
\begin{align}
    \left\langle \sigma_{n\to p}^{\pi^{+}}v\right\rangle & =1.7\ \text{mb}=4.4\times10^{-18}\ \text{eV}^{-2}\thinspace, \label{eq:-26}\\
    \left\langle \sigma_{p\to n}^{\pi^{-}}v\right\rangle /C_{\pi} & =1.5\ \text{mb}=3.9\times10^{-18}\ \text{eV}^{-2}\thinspace. \label{eq:-27}
\end{align}
Due to the opposite charges of the initial particles in $\pi^{-}+p\to n+\pi^{0}/\gamma$,
Equation \eqref{eq:-27} contains a Sommerfeld enhancement factor: $C_{\pi}=\epsilon/(1-e^{-\epsilon})$ with $\epsilon=2\pi\alpha\sqrt{\frac{m_{\pi}m_{p}}{2T(m_{\pi}+m_{p})}}$, where $\alpha$ is the fine-structure constant.
We note here that similar results can also be found in Refs.~\cite{Pospelov:2010cw} and \cite{Boyarsky:2020dzc}, despite a minus sign typo in Eq.~(D3) of Ref.~\cite{Boyarsky:2020dzc}.

\section{Constraints from \texorpdfstring{$N_{{\rm eff}}^{{\rm CMB}}$ and $Y_P$}{Neff(CMB) and YP} separately}\label{app:cons-Neff-Yp}

As discussed in the main text, HNL decays to dark sector particles affect both $N_{{\rm eff}}^{{\rm CMB}}$ and $Y_P$, and therefore, can be separately constrained by these observables.
This is presented in Fig.~\ref{fig:Neff-Yp}, where we plot the values of $\Delta N_{{\rm eff}}^{{\rm CMB}}$ and $\Delta Y_{P}$ in left and right columns, respectively.
The figure shows that $Y_P$ is more restrictive for small ${\rm Br}_X$, while $N_{{\rm eff}}^{{\rm CMB}}$ is more restrictive for large ${\rm Br}_X$.
A remarkable feature of the $N_{{\rm eff}}^{{\rm CMB}}$ constraint is that as ${\rm Br}_X$ increases, it even starts ruling out larger $U_e^2$ regions (e.g., $m_N\sim 2\ {\rm GeV}$, $|U_e|^2\sim 10^{-6}$), in addition to the small $U_e^2$ region as in the $N\to {\rm SM}$ case.
This is because a large $U_e$ together with a large $\Gamma_{N\to X}$ allows $N$ to produce $X$ while still being relativistic and in thermal equilibrium.
When $\Gamma_{N\to X}$ is sufficiently large, $X$ reaches thermal equilibrium before $N$ decouples.
Consequently, the contribution of $X$ to $N_{\rm eff}^{\rm CMB}$ becomes significant, even for short-lived $N$.
Similarly, the right panels show that this region is also disfavored by $Y_P$ since the effect of $X$ on the background evolution during BBN becomes non-negligible.
Another notable feature is that the bottom left corners of $Y_P$ plots show relatively small values of $\Delta Y_P$, implying that it is in the freeze-in regime.
Therefore, the BBN bound becomes weaker as $U_e$ decreases in this region.
This roughly corresponds to the lower edge of the BBN bound shown in Ref.~\cite{Chen:2024cla}.
The combined constraint from $N_{{\rm eff}}^{{\rm CMB}}$ and $Y_P$ measurements is shown in Fig.~\ref{fig:chi2}.

\bibliographystyle{utphys}
\bibliography{ref.bib}

@book{Dodelson:2020bqr,
    author = "Dodelson, Scott and Schmidt, Fabian",
    title = "{Modern Cosmology}",
    doi = "10.1016/C2017-0-01943-2",
    publisher = "Academic Press",
    year = "2020"
}

@article{Chen:2024cla,
    author = "Chen, Yu-Ming and Zhang, Yue",
    title = "{BBN constraint on heavy neutrino production and decay}",
    eprint = "2410.07343",
    archivePrefix = "arXiv",
    primaryClass = "hep-ph",
    doi = "10.1103/741s-211w",
    journal = "Phys. Rev. D",
    volume = "111",
    number = "12",
    pages = "123024",
    year = "2025"
}

@article{Ovchynnikov:2024xyd,
    author = "Ovchynnikov, Maksym and Syvolap, Vsevolod",
    title = "{Primordial Neutrinos and New Physics: Novel Approach to Solving the Neutrino Boltzmann Equation}",
    eprint = "2409.15129",
    archivePrefix = "arXiv",
    primaryClass = "hep-ph",
    reportNumber = "CERN-TH-2024-159",
    doi = "10.1103/PhysRevLett.134.101003",
    journal = "Phys. Rev. Lett.",
    volume = "134",
    number = "10",
    pages = "101003",
    year = "2025"
}

@article{Akita:2024ork,
    author = "Akita, Kensuke and Baur, Gideon and Ovchynnikov, Maksym and Schwetz, Thomas and Syvolap, Vsevolod",
    title = "{Dynamics of metastable standard model particles from long-lived particle decays in the MeV primordial plasma}",
    eprint = "2411.00931",
    archivePrefix = "arXiv",
    primaryClass = "hep-ph",
    reportNumber = "CERN-TH-2024-189, CERN-TH-2024-189",
    doi = "10.1103/PhysRevD.111.063542",
    journal = "Phys. Rev. D",
    volume = "111",
    number = "6",
    pages = "063542",
    year = "2025"
}

@article{Domcke:2020ety,
    author = "Domcke, Valerie and Drewes, Marco and Hufnagel, Marco and Lucente, Michele",
    title = "{MeV-scale Seesaw and Leptogenesis}",
    eprint = "2009.11678",
    archivePrefix = "arXiv",
    primaryClass = "hep-ph",
    reportNumber = "CERN-TH-2020-158, DESY-20-159, DESY 20-159",
    doi = "10.1007/JHEP01(2021)200",
    journal = "JHEP",
    volume = "01",
    pages = "200",
    year = "2021"
}

@article{Kohri:2001jx,
    author = "Kohri, Kazunori",
    title = "{Primordial nucleosynthesis and hadronic decay of a massive particle with a relatively short lifetime}",
    eprint = "astro-ph/0103411",
    archivePrefix = "arXiv",
    reportNumber = "YITP-01-23",
    doi = "10.1103/PhysRevD.64.043515",
    journal = "Phys. Rev. D",
    volume = "64",
    pages = "043515",
    year = "2001"
}

@article{Pospelov:2010cw,
    author = "Pospelov, Maxim and Pradler, Josef",
    title = "{Metastable GeV-scale particles as a solution to the cosmological lithium problem}",
    eprint = "1006.4172",
    archivePrefix = "arXiv",
    primaryClass = "hep-ph",
    doi = "10.1103/PhysRevD.82.103514",
    journal = "Phys. Rev. D",
    volume = "82",
    pages = "103514",
    year = "2010"
}

@article{Fuller:2011qy,
    author = "Fuller, George M. and Kishimoto, Chad T. and Kusenko, Alexander",
    title = "{Heavy sterile neutrinos, entropy and relativistic energy production, and the relic neutrino background}",
    eprint = "1110.6479",
    archivePrefix = "arXiv",
    primaryClass = "astro-ph.CO",
    month = "10",
    year = "2011"
}

@article{Deppisch:2024izn,
    author = "Deppisch, Frank F. and Gonzalo, Tom{\'a}s E. and Majumdar, Chayan and Zhang, Zhong",
    title = "{Relaxing limits from Big Bang Nucleosynthesis on Heavy Neutral Leptons with axion-like particles}",
    eprint = "2410.06970",
    archivePrefix = "arXiv",
    primaryClass = "hep-ph",
    reportNumber = "TTP24-038",
    doi = "10.1088/1475-7516/2025/02/054",
    journal = "JCAP",
    volume = "02",
    pages = "054",
    year = "2025"
}

@article{Kersten:2007vk,
    author = {Kersten, J{\"o}rn and Smirnov, Alexei Yu.},
    title = "{Right-Handed Neutrinos at CERN LHC and the Mechanism of Neutrino Mass Generation}",
    eprint = "0705.3221",
    archivePrefix = "arXiv",
    primaryClass = "hep-ph",
    doi = "10.1103/PhysRevD.76.073005",
    journal = "Phys. Rev. D",
    volume = "76",
    pages = "073005",
    year = "2007"
}

@article{Valerdi:2019beb,
    author = "Valerdi, Mabel and Peimbert, Antonio and Peimbert, Manuel and Sixtos, Andr{\'e}s",
    title = "{Determination of the Primordial Helium Abundance Based on NGC 346, an H ii Region of the Small Magellanic Cloud}",
    eprint = "1904.01594",
    archivePrefix = "arXiv",
    primaryClass = "astro-ph.GA",
    doi = "10.3847/1538-4357/ab14e4",
    journal = "Astrophys. J.",
    volume = "876",
    number = "2",
    pages = "98",
    year = "2019"
}

@article{Berlin:2019pbq,
    author = "Berlin, Asher and Blinov, Nikita and Li, Shirley Weishi",
    title = "{Dark Sector Equilibration During Nucleosynthesis}",
    eprint = "1904.04256",
    archivePrefix = "arXiv",
    primaryClass = "hep-ph",
    reportNumber = "FERMILAB-PUB-19-138-A-T, SLAC-PUB-17421",
    doi = "10.1103/PhysRevD.100.015038",
    journal = "Phys. Rev. D",
    volume = "100",
    number = "1",
    pages = "015038",
    year = "2019"
}

@article{Depta:2020zbh,
    author = "Depta, Paul Frederik and Hufnagel, Marco and Schmidt-Hoberg, Kai",
    title = "{Updated BBN constraints on electromagnetic decays of MeV-scale particles}",
    eprint = "2011.06519",
    archivePrefix = "arXiv",
    primaryClass = "hep-ph",
    reportNumber = "DESY-20-160, DESY 20-160, ULB-TH/20-15",
    doi = "10.1088/1475-7516/2021/04/011",
    journal = "JCAP",
    volume = "04",
    pages = "011",
    year = "2021"
}

@article{Hufnagel:2018bjp,
    author = "Hufnagel, Marco and Schmidt-Hoberg, Kai and Wild, Sebastian",
    title = "{BBN constraints on MeV-scale dark sectors. Part II. Electromagnetic decays}",
    eprint = "1808.09324",
    archivePrefix = "arXiv",
    primaryClass = "hep-ph",
    reportNumber = "DESY 18-133, DESY-18-133",
    doi = "10.1088/1475-7516/2018/11/032",
    journal = "JCAP",
    volume = "11",
    pages = "032",
    year = "2018"
}

@article{Hufnagel:2017dgo,
    author = "Hufnagel, Marco and Schmidt-Hoberg, Kai and Wild, Sebastian",
    title = "{BBN constraints on MeV-scale dark sectors. Part I. Sterile decays}",
    eprint = "1712.03972",
    archivePrefix = "arXiv",
    primaryClass = "hep-ph",
    reportNumber = "DESY-17-211",
    doi = "10.1088/1475-7516/2018/02/044",
    journal = "JCAP",
    volume = "02",
    pages = "044",
    year = "2018"
}

@article{Hambye:2021moy,
    author = "Hambye, Thomas and Hufnagel, Marco and Lucca, Matteo",
    title = "{Cosmological constraints on the decay of heavy relics into neutrinos}",
    eprint = "2112.09137",
    archivePrefix = "arXiv",
    primaryClass = "hep-ph",
    reportNumber = "ULB-TH/21-21",
    doi = "10.1088/1475-7516/2022/05/033",
    journal = "JCAP",
    volume = "05",
    number = "05",
    pages = "033",
    year = "2022"
}

@article{Abazajian:2023reo,
    author = "Abazajian, Kevork N. and Escudero, Helena Garc{\'\i}a",
    title = "{Visible in the laboratory and invisible in cosmology: Decaying sterile neutrinos}",
    eprint = "2309.11492",
    archivePrefix = "arXiv",
    primaryClass = "hep-ph",
    reportNumber = "UCI-HEP-TR-2023-09",
    doi = "10.1103/PhysRevD.108.123036",
    journal = "Phys. Rev. D",
    volume = "108",
    number = "12",
    pages = "123036",
    year = "2023"
}

@article{Alonso-Alvarez:2022uxp,
    author = "Alonso-{\'A}lvarez, Gonzalo and Cline, James M.",
    title = "{Sterile neutrino production at small mixing in the early universe}",
    eprint = "2204.04224",
    archivePrefix = "arXiv",
    primaryClass = "hep-ph",
    doi = "10.1016/j.physletb.2022.137278",
    journal = "Phys. Lett. B",
    volume = "833",
    pages = "137278",
    year = "2022"
}

@article{Aver:2015iza,
    author = "Aver, Erik and Olive, Keith A. and Skillman, Evan D.",
    title = "{The effects of He I {\ensuremath{\lambda}}10830 on helium abundance determinations}",
    eprint = "1503.08146",
    archivePrefix = "arXiv",
    primaryClass = "astro-ph.CO",
    doi = "10.1088/1475-7516/2015/07/011",
    journal = "JCAP",
    volume = "07",
    pages = "011",
    year = "2015"
}

@article{Izotov:2014fga,
    author = "Izotov, Y. I. and Thuan, T. X. and Guseva, N. G.",
    title = "{A new determination of the primordial He abundance using the He i $\lambda$10830 {\r{A}} emission line: cosmological implications}",
    eprint = "1408.6953",
    archivePrefix = "arXiv",
    primaryClass = "astro-ph.CO",
    doi = "10.1093/mnras/stu1771",
    journal = "Mon. Not. Roy. Astron. Soc.",
    volume = "445",
    number = "1",
    pages = "778--793",
    year = "2014"
}

@article{Planck:2018vyg,
    author = "Aghanim, N. and others",
    collaboration = "Planck",
    title = "{Planck 2018 results. VI. Cosmological parameters}",
    eprint = "1807.06209",
    archivePrefix = "arXiv",
    primaryClass = "astro-ph.CO",
    doi = "10.1051/0004-6361/201833910",
    journal = "Astron. Astrophys.",
    volume = "641",
    pages = "A6",
    year = "2020"
}

@article{ParticleDataGroup:2024cfk,
    author = "Navas, S. and others",
    collaboration = "Particle Data Group",
    title = "{Review of particle physics}",
    doi = "10.1103/PhysRevD.110.030001",
    journal = "Phys. Rev. D",
    volume = "110",
    number = "3",
    pages = "030001",
    year = "2024"
}

@book{Baumann:2022mni,
    author = "Baumann, Daniel",
    title = "{Cosmology}",
    doi = "10.1017/9781108937092",
    isbn = "978-1-108-93709-2, 978-1-108-83807-8",
    publisher = "Cambridge University Press",
    year = "2022"
}

@article{Meador-Woodruff:2024due,
    author = "Meador-Woodruff, Aidan and Huterer, Dragan",
    title = "{BBN-simple: How to bake a universe-sized cake}",
    eprint = "2412.07893",
    archivePrefix = "arXiv",
    primaryClass = "astro-ph.CO",
    doi = "10.1016/j.newar.2025.101722",
    journal = "New Astron. Rev.",
    volume = "100",
    pages = "101722",
    year = "2025"
}

@article{Bondarenko:2018ptm,
    author = "Bondarenko, Kyrylo and Boyarsky, Alexey and Gorbunov, Dmitry and Ruchayskiy, Oleg",
    title = "{Phenomenology of GeV-scale Heavy Neutral Leptons}",
    eprint = "1805.08567",
    archivePrefix = "arXiv",
    primaryClass = "hep-ph",
    doi = "10.1007/JHEP11(2018)032",
    journal = "JHEP",
    volume = "11",
    pages = "032",
    year = "2018"
}

@article{Sabti:2020yrt,
    author = "Sabti, Nashwan and Magalich, Andrii and Filimonova, Anastasiia",
    title = "{An Extended Analysis of Heavy Neutral Leptons during Big Bang Nucleosynthesis}",
    eprint = "2006.07387",
    archivePrefix = "arXiv",
    primaryClass = "hep-ph",
    reportNumber = "KCL-2020-09",
    doi = "10.1088/1475-7516/2020/11/056",
    journal = "JCAP",
    volume = "11",
    pages = "056",
    year = "2020"
}

@article{Bolton:2019pcu,
    author = "Bolton, Patrick D. and Deppisch, Frank F. and Dev, P. S. Bhupal",
    title = "{Neutrinoless double beta decay versus other probes of heavy sterile neutrinos}",
    eprint = "1912.03058",
    archivePrefix = "arXiv",
    primaryClass = "hep-ph",
    doi = "10.1007/JHEP03(2020)170",
    journal = "JHEP",
    volume = "03",
    pages = "170",
    year = "2020"
}

@article{Wu:2024uxa,
    author = "Wu, Quan-feng and Xu, Xun-Jie",
    title = "{High-energy and ultra-high-energy neutrinos from Primordial Black Holes}",
    eprint = "2409.09468",
    archivePrefix = "arXiv",
    primaryClass = "hep-ph",
    doi = "10.1088/1475-7516/2025/02/059",
    journal = "JCAP",
    volume = "02",
    pages = "059",
    year = "2025"
}

@article{Boyarsky:2020dzc,
    author = "Boyarsky, Alexey and Ovchynnikov, Maksym and Ruchayskiy, Oleg and Syvolap, Vsevolod",
    title = "{Improved big bang nucleosynthesis constraints on heavy neutral leptons}",
    eprint = "2008.00749",
    archivePrefix = "arXiv",
    primaryClass = "hep-ph",
    doi = "10.1103/PhysRevD.104.023517",
    journal = "Phys. Rev. D",
    volume = "104",
    number = "2",
    pages = "023517",
    year = "2021"
}

@article{Dasgupta:2013zpn,
    author = "Dasgupta, Basudeb and Kopp, Joachim",
    title = "{Cosmologically Safe eV-Scale Sterile Neutrinos and Improved Dark Matter Structure}",
    eprint = "1310.6337",
    archivePrefix = "arXiv",
    primaryClass = "hep-ph",
    doi = "10.1103/PhysRevLett.112.031803",
    journal = "Phys. Rev. Lett.",
    volume = "112",
    number = "3",
    pages = "031803",
    year = "2014"
}

@article{Chu:2018gxk,
    author = "Chu, Xiaoyong and Dasgupta, Basudeb and Dentler, Mona and Kopp, Joachim and Saviano, Ninetta",
    title = "{Sterile neutrinos with secret interactions{\textemdash}cosmological discord?}",
    eprint = "1806.10629",
    archivePrefix = "arXiv",
    primaryClass = "hep-ph",
    reportNumber = "TIFR/TH/18-17",
    doi = "10.1088/1475-7516/2018/11/049",
    journal = "JCAP",
    volume = "11",
    pages = "049",
    year = "2018"
}

@article{Chu:2015ipa,
    author = "Chu, Xiaoyong and Dasgupta, Basudeb and Kopp, Joachim",
    title = "{Sterile neutrinos with secret interactions{\textemdash}lasting friendship with cosmology}",
    eprint = "1505.02795",
    archivePrefix = "arXiv",
    primaryClass = "hep-ph",
    reportNumber = "MITP-15-033, TIFR-TH-15-14, MITP/15-033, TIFR/TH/15-14",
    doi = "10.1088/1475-7516/2015/10/011",
    journal = "JCAP",
    volume = "10",
    pages = "011",
    year = "2015"
}

@article{Hannestad:2013ana,
    author = "Hannestad, Steen and Hansen, Rasmus Sloth and Tram, Thomas",
    title = "{How Self-Interactions can Reconcile Sterile Neutrinos with Cosmology}",
    eprint = "1310.5926",
    archivePrefix = "arXiv",
    primaryClass = "astro-ph.CO",
    doi = "10.1103/PhysRevLett.112.031802",
    journal = "Phys. Rev. Lett.",
    volume = "112",
    number = "3",
    pages = "031802",
    year = "2014"
}

@article{Fuwa:2024cdf,
    author = "Fuwa, Y. and others",
    title = "{Improved measurements of neutron lifetime with cold neutron beam at J-PARC}",
    eprint = "2412.19519",
    archivePrefix = "arXiv",
    primaryClass = "nucl-ex",
    month = "12",
    year = "2024"
}

@article{Yanagisawa:2025mgx,
    author = "Yanagisawa, Hiroto and others",
    title = "{EMPRESS. XV. A New Determination of the Primordial Helium Abundance Suggesting a Moderately Low $Y_\mathrm{P}$ Value}",
    eprint = "2506.24050",
    archivePrefix = "arXiv",
    primaryClass = "astro-ph.GA",
    month = "6",
    year = "2025"
}

@article{Hasegawa:2020ctq,
    author = "Hasegawa, Takuya and Hiroshima, Nagisa and Kohri, Kazunori and Hansen, Rasmus S. L. and Tram, Thomas and Hannestad, Steen",
    title = "{MeV-scale reheating temperature and cosmological production of light sterile neutrinos}",
    eprint = "2003.13302",
    archivePrefix = "arXiv",
    primaryClass = "hep-ph",
    reportNumber = "RIKEN-iTHEMS-Report-20, UT-HET 132, KEK-Cosmo-251, KEK-TH-2204,
  IPMU20-0038",
    doi = "10.1088/1475-7516/2020/08/015",
    journal = "JCAP",
    volume = "08",
    pages = "015",
    year = "2020"
}

@article{Gelmini:2004ah,
    author = "Gelmini, Graciela and Palomares-Ruiz, Sergio and Pascoli, Silvia",
    title = "{Low reheating temperature and the visible sterile neutrino}",
    eprint = "astro-ph/0403323",
    archivePrefix = "arXiv",
    reportNumber = "UCLA-04-TEP-5",
    doi = "10.1103/PhysRevLett.93.081302",
    journal = "Phys. Rev. Lett.",
    volume = "93",
    pages = "081302",
    year = "2004"
}

@article{Barbieri:2025moq,
    author = "Barbieri, Nicola and Brinckmann, Thejs and Gariazzo, Stefano and Lattanzi, Massimiliano and Pastor, Sergio and Pisanti, Ofelia",
    title = "{Current constraints on cosmological scenarios with very low reheating temperatures}",
    eprint = "2501.01369",
    archivePrefix = "arXiv",
    primaryClass = "astro-ph.CO",
    month = "1",
    year = "2025"
}

@article{Gelmini:2008fq,
    author = "Gelmini, Graciela and Osoba, Efunwande and Palomares-Ruiz, Sergio and Pascoli, Silvia",
    title = "{MeV sterile neutrinos in low reheating temperature cosmological scenarios}",
    eprint = "0803.2735",
    archivePrefix = "arXiv",
    primaryClass = "astro-ph",
    doi = "10.1088/1475-7516/2008/10/029",
    journal = "JCAP",
    volume = "10",
    pages = "029",
    year = "2008"
}

@article{Gelmini:2019wfp,
    author = "Gelmini, Graciela B. and Lu, Philip and Takhistov, Volodymyr",
    title = "{Cosmological Dependence of Non-resonantly Produced Sterile Neutrinos}",
    eprint = "1909.13328",
    archivePrefix = "arXiv",
    primaryClass = "hep-ph",
    doi = "10.1088/1475-7516/2019/12/047",
    journal = "JCAP",
    volume = "12",
    pages = "047",
    year = "2019"
}

@article{Li:2023puz,
    author = "Li, Shao-Ping and Xu, Xun-Jie",
    title = "{N$_{eff}$ constraints on light mediators coupled to neutrinos: the dilution-resistant effect}",
    eprint = "2307.13967",
    archivePrefix = "arXiv",
    primaryClass = "hep-ph",
    doi = "10.1007/JHEP10(2023)012",
    journal = "JHEP",
    volume = "10",
    pages = "012",
    year = "2023"
}

@article{Barman:2022scg,
 archiveprefix = {arXiv},
 author = {Barman, Basabendu and Dev, P. S. Bhupal and Ghoshal, Anish},
 doi = {10.1103/PhysRevD.108.035037},
 eprint = {2210.07739},
 journal = {Phys. Rev. D},
 number = {3},
 pages = {035037},
 primaryclass = {hep-ph},
 reportnumber = {PI/UAN-2022-718FT},
 title = {{Probing freeze-in dark matter via heavy neutrino portal}},
 volume = {108},
 year = {2023}
}

@article{Chauhan:2020mgv,
 archiveprefix = {arXiv},
 author = {Chauhan, Garv and Xu, Xun-Jie},
 doi = {10.1007/JHEP04(2021)003},
 eprint = {2012.09980},
 journal = {JHEP},
 pages = {003},
 primaryclass = {hep-ph},
 title = {{How dark is the $\nu_R$-philic dark photon?}},
 volume = {04},
 year = {2021}
}

@article{Ruchayskiy:2012si,
    author = "Ruchayskiy, Oleg and Ivashko, Artem",
    title = "{Restrictions on the lifetime of sterile neutrinos from primordial nucleosynthesis}",
    eprint = "1202.2841",
    archivePrefix = "arXiv",
    primaryClass = "hep-ph",
    reportNumber = "CERN-PH-TH-2012-042",
    doi = "10.1088/1475-7516/2012/10/014",
    journal = "JCAP",
    volume = "10",
    pages = "014",
    year = "2012"
}

@article{Dolgov:2000jw,
    author = "Dolgov, A. D. and Hansen, S. H. and Raffelt, G. and Semikoz, D. V.",
    title = "{Heavy sterile neutrinos: Bounds from big bang nucleosynthesis and SN1987A}",
    eprint = "hep-ph/0008138",
    archivePrefix = "arXiv",
    doi = "10.1016/S0550-3213(00)00566-6",
    journal = "Nucl. Phys. B",
    volume = "590",
    pages = "562--574",
    year = "2000"
}

@article{Li:2022bpp,
 archiveprefix = {arXiv},
 author = {Li, Shao-Ping and Xu, Xun-Jie},
 doi = {10.1088/1475-7516/2023/06/047},
 eprint = {2212.09109},
 journal = {JCAP},
 pages = {047},
 primaryclass = {hep-ph},
 title = {{Dark matter produced from right-handed neutrinos}},
 volume = {06},
 year = {2023}
}

@article{Berryman:2019dme,
    author = "Berryman, Jeffrey M. and de Gouvea, Andre and Fox, Patrick J and Kayser, Boris Jules and Kelly, Kevin James and Raaf, Jennifer Lynne",
    title = "{Searches for Decays of New Particles in the DUNE Multi-Purpose Near Detector}",
    eprint = "1912.07622",
    archivePrefix = "arXiv",
    primaryClass = "hep-ph",
    reportNumber = "FERMILAB-PUB-19-607-ND-T, NUHEP-TH/19-16",
    doi = "10.1007/JHEP02(2020)174",
    journal = "JHEP",
    volume = "02",
    pages = "174",
    year = "2020"
}

@article{Alekhin:2015byh,
    author = "Alekhin, Sergey and others",
    title = "{A facility to Search for Hidden Particles at the CERN SPS: the SHiP physics case}",
    eprint = "1504.04855",
    archivePrefix = "arXiv",
    primaryClass = "hep-ph",
    reportNumber = "CERN-SPSC-2015-017, SPSC-P-350-ADD-1",
    doi = "10.1088/0034-4885/79/12/124201",
    journal = "Rept. Prog. Phys.",
    volume = "79",
    number = "12",
    pages = "124201",
    year = "2016"
}

@article{SHiP:2018xqw,
    author = "Ahdida, C. and others",
    collaboration = "SHiP",
    title = "{Sensitivity of the SHiP experiment to Heavy Neutral Leptons}",
    eprint = "1811.00930",
    archivePrefix = "arXiv",
    primaryClass = "hep-ph",
    doi = "10.1007/JHEP04(2019)077",
    journal = "JHEP",
    volume = "04",
    pages = "077",
    year = "2019"
}

@article{PIONEER:2022yag,
    author = "Altmannshofer, W. and others",
    collaboration = "PIONEER",
    title = "{PIONEER: Studies of Rare Pion Decays}",
    eprint = "2203.01981",
    archivePrefix = "arXiv",
    primaryClass = "hep-ex",
    month = "3",
    year = "2022"
}

@article{Krasnov:2019kdc,
    author = "Krasnov, Igor",
    title = "{DUNE prospects in the search for sterile neutrinos}",
    eprint = "1902.06099",
    archivePrefix = "arXiv",
    primaryClass = "hep-ph",
    doi = "10.1103/PhysRevD.100.075023",
    journal = "Phys. Rev. D",
    volume = "100",
    number = "7",
    pages = "075023",
    year = "2019"
}

@article{Ballett:2019bgd,
    author = "Ballett, Peter and Boschi, Tommaso and Pascoli, Silvia",
    title = "{Heavy Neutral Leptons from low-scale seesaws at the DUNE Near Detector}",
    eprint = "1905.00284",
    archivePrefix = "arXiv",
    primaryClass = "hep-ph",
    reportNumber = "IPPP/18/76",
    doi = "10.1007/JHEP03(2020)111",
    journal = "JHEP",
    volume = "03",
    pages = "111",
    year = "2020"
}

@article{Minkowski:1977sc,
    author = "Minkowski, Peter",
    title = "{$\mu \to e\gamma$ at a Rate of One Out of $10^{9}$ Muon Decays?}",
    reportNumber = "Print-77-0182 (BERN)",
    doi = "10.1016/0370-2693(77)90435-X",
    journal = "Phys. Lett. B",
    volume = "67",
    pages = "421--428",
    year = "1977"
}

@article{Mohapatra:1979ia,
    author = "Mohapatra, Rabindra N. and Senjanovic, Goran",
    title = "{Neutrino Mass and Spontaneous Parity Nonconservation}",
    reportNumber = "MDDP-TR-80-060, MDDP-PP-80-105, CCNY-HEP-79-10",
    doi = "10.1103/PhysRevLett.44.912",
    journal = "Phys. Rev. Lett.",
    volume = "44",
    pages = "912",
    year = "1980"
}

@article{Yanagida:1979as,
    author = "Yanagida, Tsutomu",
    editor = "Sawada, Osamu and Sugamoto, Akio",
    title = "{Horizontal gauge symmetry and masses of neutrinos}",
    reportNumber = "KEK-79-18-95",
    journal = "Conf. Proc. C",
    volume = "7902131",
    pages = "95--99",
    year = "1979"
}

@article{Gell-Mann:1979vob,
    author = "Gell-Mann, Murray and Ramond, Pierre and Slansky, Richard",
    title = "{Complex Spinors and Unified Theories}",
    eprint = "1306.4669",
    archivePrefix = "arXiv",
    primaryClass = "hep-th",
    reportNumber = "PRINT-80-0576",
    journal = "Conf. Proc. C",
    volume = "790927",
    pages = "315--321",
    year = "1979"
}

@article{Asaka:2005pn,
    author = "Asaka, Takehiko and Shaposhnikov, Mikhail",
    title = "{The $\nu$MSM, dark matter and baryon asymmetry of the universe}",
    eprint = "hep-ph/0505013",
    archivePrefix = "arXiv",
    doi = "10.1016/j.physletb.2005.06.020",
    journal = "Phys. Lett. B",
    volume = "620",
    pages = "17--26",
    year = "2005"
}

@article{Boyarsky:2009ix,
    author = "Boyarsky, Alexey and Ruchayskiy, Oleg and Shaposhnikov, Mikhail",
    title = "{The Role of sterile neutrinos in cosmology and astrophysics}",
    eprint = "0901.0011",
    archivePrefix = "arXiv",
    primaryClass = "hep-ph",
    doi = "10.1146/annurev.nucl.010909.083654",
    journal = "Ann. Rev. Nucl. Part. Sci.",
    volume = "59",
    pages = "191--214",
    year = "2009"
}

@article{Abdullahi:2022jlv,
    author = "Abdullahi, Asli M. and others",
    title = "{The present and future status of heavy neutral leptons}",
    eprint = "2203.08039",
    archivePrefix = "arXiv",
    primaryClass = "hep-ph",
    reportNumber = "FERMILAB-CONF-22-184-T-V",
    doi = "10.1088/1361-6471/ac98f9",
    journal = "J. Phys. G",
    volume = "50",
    number = "2",
    pages = "020501",
    year = "2023"
}

@article{He:2009ua,
    author = "He, Xiao-Gang and Oh, Sechul and Tandean, Jusak and Wen, Chung-Cheng",
    title = "{Large Mixing of Light and Heavy Neutrinos in Seesaw Models and the LHC}",
    eprint = "0907.1607",
    archivePrefix = "arXiv",
    primaryClass = "hep-ph",
    doi = "10.1103/PhysRevD.80.073012",
    journal = "Phys. Rev. D",
    volume = "80",
    pages = "073012",
    year = "2009"
}

@article{Adhikari:2010yt,
    author = "Adhikari, Rathin and Raychaudhuri, Amitava",
    title = "{Light neutrinos from massless texture and below TeV seesaw scale}",
    eprint = "1004.5111",
    archivePrefix = "arXiv",
    primaryClass = "hep-ph",
    doi = "10.1103/PhysRevD.84.033002",
    journal = "Phys. Rev. D",
    volume = "84",
    pages = "033002",
    year = "2011"
}

@article{Ibarra:2010xw,
    author = "Ibarra, A. and Molinaro, E. and Petcov, S. T.",
    title = "{TeV Scale See-Saw Mechanisms of Neutrino Mass Generation, the Majorana Nature of the Heavy Singlet Neutrinos and $(\beta\beta)_{0\nu}$-Decay}",
    eprint = "1007.2378",
    archivePrefix = "arXiv",
    primaryClass = "hep-ph",
    reportNumber = "REF-SISSA-36-2010-EP, REF-TUM-HEP-763-10, REF-IPPP-10-42, DCTP-10-84",
    doi = "10.1007/JHEP09(2010)108",
    journal = "JHEP",
    volume = "09",
    pages = "108",
    year = "2010"
}

@article{Mitra:2011qr,
    author = "Mitra, Manimala and Senjanovic, Goran and Vissani, Francesco",
    title = "{Neutrinoless Double Beta Decay and Heavy Sterile Neutrinos}",
    eprint = "1108.0004",
    archivePrefix = "arXiv",
    primaryClass = "hep-ph",
    doi = "10.1016/j.nuclphysb.2011.10.035",
    journal = "Nucl. Phys. B",
    volume = "856",
    pages = "26--73",
    year = "2012"
}

@article{Lee:2013htl,
    author = "Lee, Chang-Hun and Dev, P. S. Bhupal and Mohapatra, R. N.",
    title = "{Natural TeV-scale left-right seesaw mechanism for neutrinos and experimental tests}",
    eprint = "1309.0774",
    archivePrefix = "arXiv",
    primaryClass = "hep-ph",
    reportNumber = "UMD-PP-013-011, MAN-HEP-2013-21",
    doi = "10.1103/PhysRevD.88.093010",
    journal = "Phys. Rev. D",
    volume = "88",
    number = "9",
    pages = "093010",
    year = "2013"
}

@article{Mohapatra:1986bd,
    author = "Mohapatra, R. N. and Valle, J. W. F.",
    title = "{Neutrino Mass and Baryon Number Nonconservation in Superstring Models}",
    reportNumber = "MdDP-PP-86-127",
    doi = "10.1103/PhysRevD.34.1642",
    journal = "Phys. Rev. D",
    volume = "34",
    pages = "1642",
    year = "1986"
}

@article{Malinsky:2005bi,
    author = "Malinsky, Michal and Romao, J. C. and Valle, J. W. F.",
    title = "{Novel supersymmetric SO(10) seesaw mechanism}",
    eprint = "hep-ph/0506296",
    archivePrefix = "arXiv",
    reportNumber = "IFIC-05-28",
    doi = "10.1103/PhysRevLett.95.161801",
    journal = "Phys. Rev. Lett.",
    volume = "95",
    pages = "161801",
    year = "2005"
}

@article{Dev:2012sg,
    author = "Dev, P. S. Bhupal and Pilaftsis, Apostolos",
    title = "{Minimal Radiative Neutrino Mass Mechanism for Inverse Seesaw Models}",
    eprint = "1209.4051",
    archivePrefix = "arXiv",
    primaryClass = "hep-ph",
    reportNumber = "MAN-HEP-2012-14",
    doi = "10.1103/PhysRevD.86.113001",
    journal = "Phys. Rev. D",
    volume = "86",
    pages = "113001",
    year = "2012"
}

@article{Fernandez-Martinez:2023phj,
    author = "Fern{\'a}ndez-Mart{\'\i}nez, Enrique and Gonz{\'a}lez-L{\'o}pez, Manuel and Hern{\'a}ndez-Garc{\'\i}a, Josu and Hostert, Matheus and L{\'o}pez-Pav{\'o}n, Jacobo",
    title = "{Effective portals to heavy neutral leptons}",
    eprint = "2304.06772",
    archivePrefix = "arXiv",
    primaryClass = "hep-ph",
    reportNumber = "FTUV-23-0303.1224, IFIC/23-09",
    doi = "10.1007/JHEP09(2023)001",
    journal = "JHEP",
    volume = "09",
    pages = "001",
    year = "2023"
}

@article{Dev:2025sah,
    author = "Dev, P. S. Bhupal and Dutta, Bhaskar and Goswami, Srubabati and Tang, Jianrong Paul and Ramachandran, Aaroodd Ujjayini",
    title = "{Opening up New Parameter Space for Sterile Neutrino Dark Matter}",
    eprint = "2505.22463",
    archivePrefix = "arXiv",
    primaryClass = "hep-ph",
    reportNumber = "MI-HET-858",
    month = "5",
    year = "2025"
}

@article{Dodelson:1993je,
    author = "Dodelson, Scott and Widrow, Lawrence M.",
    title = "{Sterile-neutrinos as dark matter}",
    eprint = "hep-ph/9303287",
    archivePrefix = "arXiv",
    reportNumber = "FERMILAB-PUB-93-057-A",
    doi = "10.1103/PhysRevLett.72.17",
    journal = "Phys. Rev. Lett.",
    volume = "72",
    pages = "17--20",
    year = "1994"
}

@article{Shi:1998km,
    author = "Shi, Xiang-Dong and Fuller, George M.",
    title = "{A New dark matter candidate: Nonthermal sterile neutrinos}",
    eprint = "astro-ph/9810076",
    archivePrefix = "arXiv",
    doi = "10.1103/PhysRevLett.82.2832",
    journal = "Phys. Rev. Lett.",
    volume = "82",
    pages = "2832--2835",
    year = "1999"
}

@article{DeGouvea:2019wpf,
    author = "De Gouv{\^e}a, Andr{\'e} and Sen, Manibrata and Tangarife, Walter and Zhang, Yue",
    title = "{Dodelson-Widrow Mechanism in the Presence of Self-Interacting Neutrinos}",
    eprint = "1910.04901",
    archivePrefix = "arXiv",
    primaryClass = "hep-ph",
    reportNumber = "NUHEP-TH/19-15, FERMILAB-PUB-19-522-T",
    doi = "10.1103/PhysRevLett.124.081802",
    journal = "Phys. Rev. Lett.",
    volume = "124",
    number = "8",
    pages = "081802",
    year = "2020"
}

@article{Astros:2023xhe,
    author = "Astros, Maria Dias and Vogl, Stefan",
    title = "{Boosting the production of sterile neutrino dark matter with self-interactions}",
    eprint = "2307.15565",
    archivePrefix = "arXiv",
    primaryClass = "hep-ph",
    doi = "10.1007/JHEP03(2024)032",
    journal = "JHEP",
    volume = "03",
    pages = "032",
    year = "2024"
}

@article{Saviano:2014esa,
    author = "Saviano, Ninetta and Pisanti, Ofelia and Mangano, Gianpiero and Mirizzi, Alessandro",
    title = "{Unveiling secret interactions among sterile neutrinos with big-bang nucleosynthesis}",
    eprint = "1409.1680",
    archivePrefix = "arXiv",
    primaryClass = "astro-ph.CO",
    reportNumber = "IPPP-14-79, DCPT-14-158, IPPP-14--79, DCPT-14--158",
    doi = "10.1103/PhysRevD.90.113009",
    journal = "Phys. Rev. D",
    volume = "90",
    number = "11",
    pages = "113009",
    year = "2014"
}

@article{Forastieri:2017oma,
    author = "Forastieri, Francesco and Lattanzi, Massimiliano and Mangano, Gianpiero and Mirizzi, Alessandro and Natoli, Paolo and Saviano, Ninetta",
    title = "{Cosmic microwave background constraints on secret interactions among sterile neutrinos}",
    eprint = "1704.00626",
    archivePrefix = "arXiv",
    primaryClass = "astro-ph.CO",
    doi = "10.1088/1475-7516/2017/07/038",
    journal = "JCAP",
    volume = "07",
    pages = "038",
    year = "2017"
}

@article{Abdallah:2024uby,
    author = "Abdallah, Waleed and Gandhi, Raj and Ghosh, Tathagata and Khan, Najimuddin and Roy, Samiran and Roy, Subhojit",
    title = "{A 17 MeV pseudoscalar and the LSND, MiniBooNE and ATOMKI anomalies}",
    eprint = "2406.07643",
    archivePrefix = "arXiv",
    primaryClass = "hep-ph",
    reportNumber = "HRI-RECAPP-2024-02",
    doi = "10.1007/JHEP10(2024)086",
    journal = "JHEP",
    volume = "10",
    pages = "086",
    year = "2024"
}

@article{Bertuzzo:2018itn,
    author = "Bertuzzo, Enrico and Jana, Sudip and Machado, Pedro A. N. and Zukanovich Funchal, Renata",
    title = "{Dark Neutrino Portal to Explain MiniBooNE excess}",
    eprint = "1807.09877",
    archivePrefix = "arXiv",
    primaryClass = "hep-ph",
    reportNumber = "FERMILAB-PUB-18-336-T, OSU-HEP-18-04",
    doi = "10.1103/PhysRevLett.121.241801",
    journal = "Phys. Rev. Lett.",
    volume = "121",
    number = "24",
    pages = "241801",
    year = "2018"
}

@article{Arguelles:2018mtc,
    author = {Arg{\"u}elles, Carlos A. and Hostert, Matheus and Tsai, Yu-Dai},
    title = "{Testing New Physics Explanations of the MiniBooNE Anomaly at Neutrino Scattering Experiments}",
    eprint = "1812.08768",
    archivePrefix = "arXiv",
    primaryClass = "hep-ph",
    reportNumber = "IPPP/18/113, FERMILAB-PUB-18-686-A-ND-PPD-T",
    doi = "10.1103/PhysRevLett.123.261801",
    journal = "Phys. Rev. Lett.",
    volume = "123",
    number = "26",
    pages = "261801",
    year = "2019"
}

@article{Akita:2024nam,
    author = "Akita, Kensuke and Baur, Gideon and Ovchynnikov, Maksym and Schwetz, Thomas and Syvolap, Vsevolod",
    title = "{New Physics Decaying into Metastable Particles: Impact on Cosmic Neutrinos}",
    eprint = "2411.00892",
    archivePrefix = "arXiv",
    primaryClass = "hep-ph",
    reportNumber = "CERN-TH-2024-188, CERN-TH-2024-188",
    doi = "10.1103/PhysRevLett.134.121001",
    journal = "Phys. Rev. Lett.",
    volume = "134",
    number = "12",
    pages = "121001",
    year = "2025"
}

@misc{website1,
    howpublished="\url{https://www.hep.ucl.ac.uk/~pbolton/}"
}

@misc{website2,
    howpublished="\url{https://github.com/mhostert/Heavy-Neutrino-Limits}"
}

@misc{Wu:2026Data,
    author = {Wu, Quan-feng},
    howpublished = {\href{https://doi.org/10.5281/zenodo.18785619}{Zenodo.18785619}},
    title = {Code for arXiv:2507.12270 (1.0.0)},
    year = {2026},
    note = {\url{https://github.com/Fenyutanchan/code-2507.12270}},
}

\end{document}